\documentclass[11pt]{article} 

\usepackage{url,hyperref,lineno,microtype,subcaption}
\usepackage[onehalfspacing]{setspace}

\usepackage{amsmath,amsfonts,amssymb,graphicx,amsthm,geometry}
\usepackage{stackrel,stmaryrd,mathabx,etoolbox,framed,accents}
\usepackage[all]{xy}

\def\be{\begin{equation}}
\def\ee{\end{equation}}
\def\ba{\begin{eqnarray}}
\def\ea{\end{eqnarray}}

\makeatletter
\newsavebox{\@brx}
\newcommand{\llangle}[1][]{\savebox{\@brx}{\(\m@th{#1\langle}\)}%
  \mathopen{\copy\@brx\kern-0.5\wd\@brx\usebox{\@brx}}}
\newcommand{\rrangle}[1][]{\savebox{\@brx}{\(\m@th{#1\rangle}\)}%
  \mathclose{\copy\@brx\kern-0.5\wd\@brx\usebox{\@brx}}}
\makeatother

\newgeometry{vmargin={25mm}, hmargin={22mm,22mm}} 

\begin{document}

\setcounter{tocdepth}{1}

\title{Notes on higher-spin diffeomorphisms} 

\author{Xavier Bekaert}

\date{Institut Denis Poisson, Unit\'e Mixte de Recherche $7013$ du CNRS\\
Universit\'e de Tours, Univrsit\'e d'Orl\'eans\\
Parc de Grandmount, 37200 Tours, France\\
\vspace{2mm}
{\tt xavier.bekaert@lmpt.univ-tours.fr}
}

\maketitle

\vspace{5mm}

\begin{abstract}

Higher-spin diffeomorphisms are to higher-order differential operators what diffeomorphisms are to vector fields.
Their rigorous definition is a challenging mathematical problem which might predate a better understanding of higher-spin symmetries and interactions. Several yes-go and no-go results on higher-spin diffeomorphisms are collected from the mathematical literature in order to propose a generalisation of the algebra of differential operators on which higher-spin diffeomorphisms are well-defined.

\vspace{5mm}

\noindent\textit{This work is dedicated to the memory of Christiane Schomblond, who taught several generations of Belgian physicists the formative rigor and delicate beauty of theoretical physics.}
\end{abstract}

\thispagestyle{empty}

\pagebreak

\setcounter{page}{1}

\tableofcontents

\section{Introduction}

Higher-spin gravity theories are interacting theories whose spectrum of free particles contains at least one massless higher-spin particle (\textit{i.e.} of spin greater than two). As a byproduct of the consistency of their symmetries, they must contain as well as a massless particle of spin two in their spectrum, which one may tentatively interpret as a graviton (hence the name ``gravity''). 
Despite the long history of this subject\footnote{Many pedagogical reviews of various levels are available by now: advanced ones \cite{Reviews} as well as introductory ones \cite{Introductions}. Two books of conference proceedings also offer a panorama of this research area \cite{Proceedings}.}, the (non)locality properties of its finite gauge symmetries remain elusive.\footnote{A distinct but related issue is the degree of (non)locality of higher-spin interactions. It has been a subject of intense scrutiny and debate over the last years. This open problem will not be adressed here, although one may hope that the unavoidable non-locality of finite higher-spin gauge symmetries might shed some light on this issue in the future.} 
They are the subject of this paper. More precisely, one will focus on finite gauge symmetries in the metric-like formulation with unconstrained symmetric tensors as gauge fields and parameters. 

It is well-known that a finite collection of symmetric tensor fields can be packaged into a single generating function, which can be interpreted as the symbol of a differential operator. Accordingly, the commutator of differential operators (or, respectively, the Poisson bracket of their symbols) defines a Lie algebra structure on the space of higher-spin gauge parameters. Moreover, the latter gauge parameters act on higher-spin gauge fields via the adjoint action. This simple procedure provides a non-abelian deformation of the gauge symmetries for free higher-spin gauge fields in the unconstrained metric-like formulation.
These non-abelian deformations have been first considered by Segal in his investigation \cite{Segal:2002gd} of conformal higher-spin gravity (where they are supplemented by higher-spin Weyl-like transformations). 
Later on, they were obtained by gauging hypertranslations via the Noether method in the metric-like formulation \cite{Bekaert:2009ud}, the frame-like formulation \cite{Bonora:2018uwx} and the BRST formulation \cite{Fotopoulos_2007}. They should also arise from off-shell higher-spin gravity after elimination of auxiliary fields and partial gauge-fixing \cite{offshellV}. Finally, they appear as the natural gauge symmetries in ambient space for the higher-spin extension of Fefferman-Graham's ambient metric \cite{HSFG}.

Some recent mathematical results by Grabowski and Poncin \cite{Grabowski:2002,Grabowski:2003b} on the automorphisms of the Lie algebra of differential operators imply that this Lie algebra does not integrate to a Lie group. This mathematical no-go theorem implies that non-abelian higher-spin gauge symmetries are not well-defined if the fields and parameters are packed into differential operators.
This shows that if finite higher-spin symmetries are taken seriously, then they require to leave the realm of operators with bounded number of derivatives (the landmark of locality). Some proposals are made in this direction in this paper, mostly by collecting old results in deformation quantisation.

\vspace{3mm}

The plan of the paper is as follows. 
The problem with the finite counterpart of non-abelian higher-spin gauge symmetries in the unconstrained metric-like formulation is addressed in Section \ref{sect2}, where the results by Grabowski and Poncin are briefly reviewed and translated into a no-go theorem on  naive higher-spin diffeomorphisms. This no-go theorem calls for an extension of the class of generating functions for higher-spin gauge fields and parameters. This way out is exemplified in Section \ref{symplectomcotgt} on the geometrically transparent case of their Poisson limit: the space of symbols may be extended to the space of smooth functions on the cotangent bundle on which symplectomophisms are well-defined. 
Our strategy is to consider a deformation of this example where the space of differential operators is extended to a larger space, modeled on the deformed algebra of functions on the cotangent bundle. This is done by importing known results on the deformation quantisation of cotangent bundles. Nevertheless, some work needs to be done in order to ensure that the elements of the corresponding deformed algebra can be interpreted as suitable generalisations of differential operators. This is the main technical task of this paper. As a preliminary step, the one-to-one correspondence(s) between differential operators and their symbols, \textit{i.e.} the quantisation(s) of the cotangent bundle, is reviewed in Section \ref{quantisations} in the light of almost-commutative algebras. In Section \ref{beyond}, our general strategy for going beyond differential operators is explained. As a first proposal, the class of almost-differential is defined in Section \ref{almostdiff} in terms of differential operators weighted by some formal variable. This provides a  functional space of generalised differential operators bypassing the no-go theorem. Some results from deformation quantisation are then reviewed in Section \ref{defoquant}.
The next step is to make use of these results to define in Section \ref{quasidiffops} the class of formal quasi-differential operators, and show in Section \ref{HSdiffsfinally} that it provides another space of generalised differential operators bypassing the no-go theorem.
The relation between the two classes is discussed in Section \ref{quotientalgalmostdiffops}. Finally, we end up with a short conclusion in Section \ref{conclusion}. An appendix details the proof of a technical lemma. The paper is long but it aims to be as self-contained as possible.

\section{Higher-spin gauge symmetries in the unconstrained metric-like formulation}\label{sect2}

\subsection{Non-abelian deformations of higher-spin gauge symmetries}

The metric-like formulation dates back to Fronsdal's seminal works on free higher-spin gauge fields on constant-curvature backgrounds \cite{Fronsdal:1978rb}. He immediately raised the question of the nonlinear completion of this free theory, \textit{i.e.} the introduction of interactions in a way compatible with (deformed) gauge symmetries. 

\paragraph{Deforming higher-spin gauge symmetries.} A more humble problem, that one can view as a preliminary step towards Fronsdal's programme, is to look for a deformation 
\be\label{deformedgtransfo}
\delta_\xi\, h_{\mu_1\cdots\mu_s}\,=\,s\,\nabla_{(\mu_1}\xi_{\mu_2\cdots\mu_s)}\,+\,\mathcal{O}(h)\,,
\ee
of the infinitesimal gauge transformations, where $\nabla$ is the covariant derivative with respect to the background metric $\bar{g}_{\mu\nu}$ with respect to which indices will be raised and lowered. The round bracket stands for the total symmetrisation with weight one (\textit{i.e.} $T_{(\mu_1\cdots\mu_s)}=T_{\mu_1\cdots\mu_s}$ for a symmetric tensor). One requires that the deformation must be consistent, such that the commutator $[\delta_{\xi_1},\delta_{\xi_2}]$ of two gauge transformations closes, 
\be
\big[\delta_{\xi_1},\delta_{\xi_2}\big] h_{\mu_1\cdots\mu_s}\,=\,\delta_{[\xi_1,\xi_2]}h_{\mu_1\cdots\mu_s}\,,
\ee
where $[\xi_1,\xi_2]$ stands for a Lie bracket over the space of gauge parameters. Strictly speaking, the closure might hold only on-shell. 

\paragraph{Non-abelian deformations.} The deformation \eqref{deformedgtransfo} is usually required to be non-abelian, \textit{i.e.} the Lie bracket over the space of gauge parameters must be non-trivial. In the original metric-like formulation of Fronsdal, this deformation problem is already a challenge\footnote{This difficulty was recognised immediately by Fronsdal, despite he actually found such a non-abelian deformation together with a Lie bracket over the space of traceless symmetric tensor fields \cite{Fronsdal:1979gk}.} because, in the undeformed theory, the gauge fields $h_{\mu_1\cdots\mu_s}$ are constrained to be double-traceless while the gauge parameters $\xi_{\mu_1\cdots\mu_{s-1}}$ are traceless, with respect to the background metric.

\paragraph{Relaxing trace constraints.} This difficulty provides a strong motivation for considering ``unconstrained'' formulations\footnote{Various metric-like unconstrained formulations of free massless higher-spin fields are available by now (see e.g. the reviews \cite{FS} and refs therein). They will not be reviewed here since our considerations are not dynamical but purely at the level of gauge symmetries.} where these tracelessness conditions are absent.\footnote{Let us repeat that, in the present paper, the focus is on finite gauge symmetries in the unconstrained metric-like formulation for technical simplicity. It is natural to expect that our main conclusions should apply to the original constrained metric-like formulation of Fronsdal without any qualitative change.}
In such case, two (closely related) non-abelian deformations stand out from the crowd, for their simplicity. They have a neat geometric interpretation which makes manifest that these fully nonlinear gauge transformations are actually background independent (although they look superficially background-dependent deformations, if one perturbs around a given constant-curvature background).
Let us briefly review these infinitesimal gauge transformations, since their finite counterpart is the focus of this paper.

\subsection{Two examples of infinitesimal higher-spin gauge symmetries}

The crucial observation is that the undeformed part of higher-spin gauge transformation in \eqref{deformedgtransfo}
takes the form of a so-called \textit{Killing derivative}, $\nabla_{(\mu_1}\xi_{\mu_2\cdots\mu_s)}$, which is familiar to geometers and well-known to arise from the Schouten bracket with the metric. This remark can be used to produce two examples of nonabelian deformations as follows.

\paragraph{Example 1 (Hamiltonian vector fields on the cotangent bundle).} First, one packages the tower of symmetric tensors (here gauge fields and parameters) into a single function on the cotangent bundle $T^*M$ of the spacetime manifold $M$\,:
\be\label{symbols}
h(x,p)\,=\,\sum\limits_{s\geqslant 0}\frac1{s!}\,h^{\mu_1\cdots\mu_s}(x)\,p_{\mu_1}\cdots p_{\mu_s}\,,\qquad\xi(x,p)\,=\,\sum\limits_{s\geqslant 1}\frac1{(s-1)!}\,\xi^{\mu_1\cdots\mu_{s-1}}(x)\,p_{\mu_1}\cdots p_{\mu_{s-1}}\,.
\ee
Note that the coefficients in the expansion in powers of momenta are contravariant symmetric tensor fields. The latter will be called \textit{symmetric multivector fields} for short.
Second, one defines the \textit{higher-spin metric} as the following extension of the background metric
\be\label{HSmetric}
g(x,p)\,=\,g_0(x,p)\,+\,h(x,p)\,,\qquad g_0(x,p)=\frac12\,\bar{g}^{\mu\nu}(x)\,p_\mu\,p_\nu
\ee
where $g_0(x,p)$ encodes the background metric, of which $h(x,p)$ is seen as a small (higher-spin) fluctuation.
Third, the canonical Poisson bracket on the cotangent bundle $T^*M$ provides a non-abelian deformation of higher-spin gauge transformations,
\be\label{classdefogtransfo}
\delta_\xi\,g(x,p)\,=\,\{\,\xi(x,p)\,,\,g(x,p)\,\}\,.
\ee
One can check that \eqref{classdefogtransfo} takes the form \eqref{deformedgtransfo} by inserting definitions \eqref{symbols}-\eqref{HSmetric} and the formula
\be
\{\,\xi(x,p)\,,\,g_0(x,p)\,\}\,=\,\sum\limits_{s\geqslant 1} \frac1{(s-1)!}\,\nabla{}^{(\mu_1}\xi^{\mu_2\cdots\mu_s)}(x)\,p_{\mu_1}\cdots p_{\mu_s}\,.
\ee
The corresponding Lie bracket over gauge parameters is nothing but the canonical Poisson bracket 
\be
\{\,\xi_1(x,p)\,,\,\xi_2(x,p)\,\}\,=\,\frac{\partial\xi_1(x,p)}{\partial x^{\mu}}\frac{\partial\xi_2(x,p)}{\partial p_{\mu}}-\frac{\partial\xi_1(x,p)}{\partial p_{\mu}}\frac{\partial\xi_2(x,p)}{\partial x^{\mu}}\,.
\ee
When expressed in terms of the coefficients in \eqref{symbols}, it is called the \textit{Schouten bracket of symmetric multivector fields} \cite{Dubois-Violette:1994tlf} and reads
\be\label{Schouten}
\{\, \xi_1\,, \xi_2\,\}^{\nu_1\cdots\nu_{r_1+r_2-1}}\,=\,r_2\,\partial_\mu  \xi_1^{\,(\nu_1\cdots\nu_{r_1}}(x)\,\xi_2^{\,\nu_{r_1+1}\cdots\nu_{r_1+r_2-1})\mu}(x)\,-\, r_1\,\partial_\mu  \xi_2^{\,(\nu_1\cdots\nu_{r_2}}(x)\,\xi_1^{\,\nu_{r_2+1}\cdots\nu_{r_1+r_2-1})\mu}(x)\,.
\ee
In symplectic geometry language, the higher-spin gauge transformation \eqref{classdefogtransfo} is nothing but a Lie derivative 
of the function $g(x,p)$ along the Hamiltonian vector field on the cotangent bundle $T^*M$ generated by the function $\xi(x,p)$. Therefore its finite counterpart is a Hamiltonian symplectomorphism of the cotangent bundle.

\paragraph{Example 2 (Higher-spin Lie derivatives).} Higher-spin gauge fields $h_{\mu_1\cdots\mu_s}$ transform as Lorentz symmetric tensor fields on a constant-curvature background, at linearised order. However, there is no reason to expect them to transform like symmetric tensor fields under general coordinate transformations in fully nonlinear higher-spin gravity. In fact, minimal coupling of massless particles to gravity is known to be problematic in the standard (metric-like or frame-like) formulations.\footnote{See e.g. the 2nd and 3rd refs in \cite{Introductions} for some reviews of no-go theorems. See also \cite{Krasnov:2021nsq} for a recent discussion of ways out in other formulations (such as lightcone or twistor).}
Relaxing this requirement, one can package the tower of gauge fields and parameters as differential operators on the spacetime manifold $M$\,:
\be\label{HX}
\hat{H}\,=\,\sum\limits_{s\geqslant 0}\frac{(-i\ell)^s}{s!}\,h^{\mu_1\cdots\mu_s}(x)\,\nabla_{\mu_1}\cdots\nabla_{\mu_s}+\,\ldots\,,\quad \hat{X}\,=\,\sum\limits_{s\geqslant 1}\frac{(-i\ell)^{s-1}}{(s-1)!}\,\xi^{\mu_1\cdots\mu_{s-1}}(x)\,\nabla_{\mu_1}\cdots\nabla_{\mu_{s-1}}+\,\ldots
\ee
where $\ell$ is a parameter with the dimension of a length (which plays a similar role to the string length) and the dots stand for lower-order terms fixed by the ordering prescription.\footnote{The prescription for \eqref{HX} is as follows: one consider a covariantised Weyl map $\mathcal{W}:h(x,p)\mapsto\hat{H}$ where the operators are obtained from their symbols \eqref{symbols} via (i) the quantisation rule $p\mapsto -i\ell\nabla$ and (ii) the anticommutator-ordering prescription with respect to the covariant derivative
\be
\hat{H}\,=\,\left.\exp\left(\-\,\frac{i}2\,\ell\,[\,\nabla_\mu\,,\,\,]_{_+}\,\frac{\partial}{\partial p_\mu}\right)h(x,p)\right|_{p=0}\,=\,
\sum\limits_{s\geqslant 0}\frac{(-i\ell)^s}{2^s\,s!}\Big[\nabla_{\mu_1}\,,\,\big[\nabla_{\mu_2}\,,\, \cdots\,,\,[\nabla_{\mu_s}\,,\,h^{\mu_1\cdots\mu_s}(x)]_{_+}\cdots\big]_+\Big]_+\,,
\ee
where $[\hat{A},\hat{B}]_+:=\hat{A}\circ\hat{B}+\hat{B}\circ\hat{A}$ stands for the anticommutator. This procedure may look complicated but this is a purely technical trick to ensure that \eqref{quantumdefogtransfo} matches the form \eqref{deformedgtransfo}.
It will not be used later on. (In any case, see Equation (3.18) and Appendix A in 
 \cite{Bekaert:2009ud} where the flat case is explained in details.)}
Similarly, the higher-spin metric defines the following extension of the background Laplacian
\be\label{HSmetric2}
\hat{G}\,=\,\hat{G}_0\,+\,\hat{H}\,,\qquad \hat{G}_0\,=\,-\,\frac{\ell^2}2\,\nabla^2\,.
\ee
The commutator of differential operators provides another non-abelian deformation \eqref{deformedgtransfo} of higher-spin gauge transformations,
\be\label{quantumdefogtransfo}
\delta_{\hat{X}}\hat{G}\,=\,\frac{i}{\ell}\,[\hat{X},\hat{G}]\,,
\ee
since 
\be
[\hat{X},\hat{G}_0]=\sum\limits_{s\geqslant 0}\frac{(-i\ell)^{s}}{(s-1)!}\nabla^{(\mu_1}\xi^{\mu_2\cdots\mu_s)}(x)\,\nabla_{\mu_1}\cdots \nabla_{\mu_s}+\ldots
\ee 

\paragraph{Low-spin truncation.} If one truncates the gauge transformations \eqref{deformedgtransfo} obtained from \eqref{classdefogtransfo} or \eqref{quantumdefogtransfo} to the ``low-spin'' sector ($s=1,2$), they both reproduce Maxwell gauge transformations of scalar gauge parameter $\xi(x)$ and infinitesimal diffeomorphims, \textit{i.e.} Lie derivatives along the vector field $X=\xi^\mu(x)\partial_\mu$. By analogy, the unconstrained gauge transformation \eqref{quantumdefogtransfo} will be called a higher-spin Lie derivative along the differential operator $\hat{X}\,$. Their finite counterpart will be called \textit{higher-spin diffeomorphisms} (they will be defined more precisely later).

\paragraph{Differential operators vs Symbols.} Note that a proper ordering prescription implies that there is a one-to-one correspondence between differential operators and symbols. In particular, here there is a one-to-one correspondence between higher-spin metrics $g(x,p)$ as in \eqref{HSmetric}
and higher-spin extension $\hat{G}$ of the Laplacian as in \eqref{HSmetric2}.
Accordingly, one can rewrite \eqref{quantumdefogtransfo} as
\be\label{quantdefogtransfoo}
\delta_\xi\,g(x,p)\,=\,\frac{i}{\ell}\,\big[\,\xi(x,p)\,,\,g(x,p)\,\big]_\star\,=\,\{\,\xi(x,p)\,,\,g(x,p)\,\}+{\mathcal O}(\ell)\,,
\ee
in terms of a suitable star-product $\star$ (with respect to $\ell$ as formal variable).

\paragraph{Higher-spin diffeomorphisms vs Hamiltonian symplectomorphisms.} The form \eqref{quantdefogtransfoo} of the transformation \eqref{quantumdefogtransfo} makes manifest that the first deformation \eqref{classdefogtransfo} can be obtained as the Poisson limit ($\ell\to 0$) of the second deformation \eqref{quantumdefogtransfo}.
Accordingly, the Poisson limit of higher-spin diffeomorphisms are symplectomorphisms of the cotangent bundle of spacetime. In this sense, higher-spin diffeomorphisms can be thought as quantum symplectomorphisms, however we refrain from using this terminology because it can be misleading (they are finite gauge symmetries of \textit{classical} higher-spin gravity).\footnote{Let us stress that the words ``quantum'', ``quantisation'', etc, throughout this paper should be taken in a mathematical technical sense, not in a physical literal sense. In deformation quantisation, ``quantum'' is synonymous of ``associative'' while ``classical'' is synonymous of ``Poisson''.}

\subsection{Problems with higher-spin diffeomorphisms}

The existence of finite counterparts of the infinitesimal non-abelian gauge transformations \eqref{classdefogtransfo} and \eqref{quantumdefogtransfo} is a challenging mathematical problem. The subtlety here is the functional space to which the higher-spin metric $g(x,p)$ (or, equivalently, $\hat{G}$) and the higher-spin gauge parameter $\xi(x,p)$ should belong in order for finite higher-spin gauge transformations to be well-defined. 

The problem is that symplectomorphisms or higher-spin diffeomorphisms generically transform a symbol $g(x,p)$ or a differential operator $\hat{G}$, that encodes a finite collection of symmetric tensor gauge fields, into 
a ``pseudo'' symbol (in the sense of a function on the cotangent bundle which is \textit{not} polynomial in the momenta) or, respectively, a ``pseudo'' differential operator of infinite order (in the sense of an operator which is \textit{not} polynomial in the derivatives). 
Generically, a finite higher-spin gauge transformation of a given higher-spin metric activates an infinite tower of higher-spin gauge fields with unbounded spin. This agrees with (and is closely related to) the standard lore that higher-spin gravity theories (in dimension 4 or higher) must have an infinite spectrum of gauge fields with unbounded spin. While the non-abelian structure of \textit{infinitesimal} higher-spin symmetries (rigid or gauged) is enough to derive the latter property of higher-spin gravity spectra, the problem with \textit{finite} higher-spin gauge symmetries is much stronger. It highlights the fact that higher-spin diffeomorphisms are necessarily non-local in terms of the spacetime manifold (albeit local in terms of its cotangent bundle).

These general remarks will not come as a surprise to experts. Our goal here is to emphasise that both the problem and some solutions can be made mathematically precise by extracting known results from the mathematical literature. 

\subsection{Notation and terminology}

In order to state the problem as a theorem, let us fix some notation and terminology.\footnote{For the sake of simplicity, reality conditions will not be discussed in this paper. All algebras considered here will be complex, of which suitable reals forms (e.g. of Hermitian operators) can be extracted if necessary. From now on, factors of $i$ will be dropped from all formulae.}

\paragraph{Poisson algebra of symbols.} Let $\mathcal{S}(M)$ denote the space of smooth functions on the cotangent bundle $T^*M$ that are polynomial in the momenta. Such functions are usually called \textit{symbols}. In Darboux coordinates, symbols $h(x,p)$ are smooth functions of positions $x^\mu$ and polynomial functions of momenta $p_\nu$\,. The space $\mathcal{S}(M)$ of symbols on $M$ is endowed with a structure of Poisson algebra via the pointwise product and the canonical Poisson bracket. It is isomorphic to the space $\odot\mathcal{T}(M):=\Gamma(\odot TM)$ of sections of the symmetric tensor product of the tangent bundle (\textit{i.e.} symmetric multivector fields on $M$), endowed with a structure of Poisson algebra via the symmetric tensor product and the Schouten bracket \eqref{Schouten}.

\paragraph{Associative algebra of differential operators.} Let $\mathcal{D}(M)$ denote the associative algebra of differential operators on $M$. Heuristically, they are linear operators acting on $C^\infty(M)$ of the form $\hat{H}(x,\partial)$ with only a finite number of derivatives.

\subsection{No-go theorems}\label{nogoths}

The Lie algebras corresponding to the vector spaces $\mathcal{S}(M)$ and $\mathcal{D}(M)$ endowed with the canonical Poisson bracket or, respectively, with the commutator as Lie bracket will be denoted 
$\mathfrak{S}(M)$ and $\mathfrak{D}(M)$.
The higher-spin gauge transformations \eqref{classdefogtransfo} and \eqref{quantumdefogtransfo} respectively correspond to the adjoint action of the respective algebras $\mathfrak{S}(M)$ and $\mathfrak{D}(M)$ on themselves. 

By definition, Hamiltonian vector fields on $T^*M$ are inner derivations of the Poisson algebra $C^\infty(T^*M)$. In particular, Hamiltonian vector fields on $T^*M$ which are polynomial in the momenta are inner derivations of the Poisson algebra $\mathcal{S}(M)$ of symbols. Similarly, the adjoint action of a differential operator is by definition an inner derivation of the associative algebra $\mathcal{D}(M)$.
Therefore, the higher-spin gauge transformations \eqref{classdefogtransfo} and \eqref{quantumdefogtransfo} coincide with the inner derivations of these two algebras. However, the inner automorphisms of these algebras are scarce: they only correspond to the gauge symmetries of the low-spin truncation, \textit{i.e.} internal Abelian gauge symmetries (as in Maxwell electromagnetism) and standard diffeomorphisms of the manifold $M$ (as in general relativity). 

The Lie algebras of inner derivations of $\mathcal{S}(M)$ and $\mathcal{D}(M)$ are isomorphic to $\mathfrak{S}(M)$ and $\mathfrak{D}(M)$, respectively. The problem is that these Lie algebras of inner derivations do not integrate to Lie groups of inner automorphisms. While the infinitesimal automorphisms are perfectly well defined, however their naive exponentiation is not well defined in general (see Section 7 of \cite{Grabowski:2002} and Section 8 of \cite{Grabowski:2003b} for more details) except for the inner automorphisms of $\mathfrak{S}(M)$ or $\mathfrak{D}(M)$ generated by symbols of degree one or, respectively, by first-order differential operators. The Poisson algebra  $\mathcal{S}(M)$ of symbols and the associative algebra $\mathcal{D}(M)$ of differential operators admit very few \textit{finite} automorphisms, although they admit a plethora of \textit{infinitesimal} automorphisms (derivations). 

One may summarise the results of \cite{Grabowski:2002,Grabowski:2003b} relevant for us as follows.

\vspace{3mm}
\noindent{\textbf{Theorem (Grabowski \& Poncin)\,:}
\textit{Any one-parameter group of automorphisms of the associative algebra ${\mathcal D}(M)$ of differential operators (respectively, of the Poisson algebra ${\mathcal S}(M)$ of symbols) is generated by a first-order differential operator (respectively, by a symbol of degree one in the momenta).}}
\vspace{3mm}

By contraposition, this can be expressed equivalently as a no-go theorem.

\vspace{3mm}
\noindent\textbf{No-go theorem\,:} \textit{Higher-spin Lie derivatives along higher-order differential operators on $M$ (respectively, higher-degree Hamiltonian vector fields on $T^*M$) cannot be integrated to one-parameter groups of automorphisms of the associative algebra ${\mathcal D}(M)$ of differential operators (respectively, of the Poisson algebra ${\mathcal S}(M)$ of symbols).}
\vspace{3mm}

If a Lie algebra is integrable to a Lie group\footnote{Note that the so-called ``Lie's third theorem'' stating that every finite-dimensional Lie algebra $\mathfrak g$ over the real numbers is associated to a Lie group $G$ does not hold in general for infinite-dimensional Lie algebras.}, then one would expect (for any reasonable topology) that all its inner derivations are integrable, locally, to one-parameter groups of automorphisms of the Lie algebra (via the exponential map).
Accordingly, an important corollary of Grabowski-Poncin's theorem is the strong no-go theorem of (\textit{cf.} Corollary 4 in \cite{Grabowski:2002})\,:

\vspace{3mm}
\noindent\textbf{Corollary (Grabowski \& Poncin)\,:} \textit{The two infinite-dimensional Lie algebras, ${\mathfrak{D}}(M)$ of differential operators and $\mathfrak{S}(M)$ of symbols on a manifold $M$, are not integrable to infinite-dimensional Lie groups (of which they are the Lie algebras).}
\vspace{3mm}

\subsection{Definitions}

In order to describe precisely the origin of the problem, some definitions are in order.

\paragraph{Filtered vs Graded algebras.} A filtered algebra ${\mathcal A}$ admits a collection of vector subspaces ${\mathcal A}_i$ (indexed by non-negative integers $i\in\mathbb N$) such that ${\mathcal A}_i\subset {\mathcal A}_j$ for $i<j$ and 
 ${\mathcal A}_i{\mathcal A}_j\subseteq {\mathcal A}_{i+j}$ for all $i$ and $j$.\footnote{Moreover, if the algebra $\mathcal A$ has a unit element, then one further requires that $1\in {\mathcal A}_0$. Here, all associative algebras are assumed to have a unit element and, accordingly, morphisms of associative algebras relate their unit elements.}  
The \textit{graded algebra associated to the filtered algebra} ${\mathcal A}$ is denoted $\text{gr}{\mathcal A}=\oplus_{i\in\mathbb N}\,\text{gr}_i{\mathcal A}$ and defined via the quotients $\text{gr}_i{\mathcal A}={\mathcal A}_i\,/{\mathcal A}_{i-1}$. The equivalence class $[a]\in \text{gr}{\mathcal A}$ of an element $a\in{\mathcal A}$ of a filtered algebra will be called the \textit{principal symbol} of $a$. The principal symbol is a homogeneous element of the associated graded algebra (\textit{i.e.} $[a]$ has a fixed grading). This defines an infinite collection of surjective linear maps 
\be\label{principsymb}
\sigma_i\,:\,{\mathcal A}_i\twoheadrightarrow\text{gr}_i{\mathcal A}\,:\,a_i\mapsto[a_i]\,,
\ee 
which will be collectively denoted $\sigma:{\mathcal A}\twoheadrightarrow\text{gr}{\mathcal A}$ and called the \textit{principal symbol map}.

\paragraph{Commutative algebra of symbols.} The algebra $\mathcal{S}(M)$ of symbols can be either filtered or graded by the polynomial degree in the momenta of symbols. The distinction only has to do with the choice of decomposition of this algebra, either as a ``matryoshka doll'' (filtered algebra) or as a ``sliced bread'' (graded algebra). A symbol on $M$ which is a homogeneous polynomial in the momenta will be called a \textit{principal symbol}. Principal symbols are in one-to-one correspondence with symmetric multivector fields, hence $\mathcal{S}(M)\cong\odot\mathcal{T}(M)$ as graded algebras.

\paragraph{Almost-commutative algebras.}
The graded algebra $\text{gr}{\mathcal A}$ associated to a filtered associative algebra $\mathcal A$ is commutative iff the commutator is of degree $-1$\,: $[{\mathcal A}_i,{\mathcal A}_j]\subseteq {\mathcal A}_{i+j-1}$. In such case, the filtered associative algebra $\mathcal A$ is called \textit{almost commutative} (because it is commutative up to lower-order terms).

\paragraph{Differential operators vs principal symbols.} The associative algebra $\mathcal{D}(M)$, filtered by the order of differential operators, is almost-commutative. 
The principal symbol of a differential operator of order $k$ is equivalent to a symmetric multivector field of rank $k$. The principal symbol map $\sigma:\mathcal{D}(M)\twoheadrightarrow\odot\mathcal{T}(M)$ stands for the collection of surjective linear maps
\be\label{prsymb}
\sigma_k\,:\,\mathcal{D}^k(M)\twoheadrightarrow\odot^k\mathcal{T}(M)\,:\,\sum\limits_{r=0}^k \frac1{r!}\,X^{\mu_1\cdots \mu_r}(x)\,\partial_{\mu_1}\cdots \partial_{\mu_r}\,\mapsto\,X^{\mu_1\cdots \mu_k}(x)\partial_{\mu_1}\odot\cdots\odot\partial_{\mu_k}\,.
\ee

\paragraph{Schouten algebras.} A Poisson algebra ${\mathcal P}$ that is a graded algebra for the commutative product (\textit{i.e.} ${\mathcal P}_i\cdot{\mathcal P}_j\subseteq {\mathcal P}_{i+j}$ for the commutative product) and whose Poisson bracket is of grading $-1$ (\textit{i.e.} $\{{\mathcal P}_i,{\mathcal P}_j\}\subseteq {\mathcal P}_{i+j-1}$) will be called a \textit{Schouten algebra}.\footnote{Such a Poisson algebra was called a ``classical Poisson algebra'' in \cite{Grabowski:2002}. Since this terminology may be confusing to people familiar with the vocabulary in deformation quantisation (where Poisson algebras are, by definition, classical), one chose to call them Schouten algebras (as a tribute to the Schouten bracket of symmetric multivector fields). Note that a Gerstenhaber algebras are the supercommutative analogues of Schouten algebras.}
The Poisson algebras $\text{gr}\,\mathcal{S}(M)$ of principal symbols and $\odot\mathcal{T}(M)$ of symmetric multivector fields, respectively graded by the polynomial degree in the momenta and by the rank, are isomorphic as Schouten algebras.
 
\paragraph{Poisson limit.} The graded algebra $\text{gr}{\mathcal A}$ of an almost-commutative algebra $\mathcal A$ is endowed with a canonical structure of Schouten algebra, where the Poisson bracket $\{\,\,,\,\}$ is inherited from the commutator bracket $[\,\,,\,]$ via the principal symbol: $\{\sigma(a),\sigma(b)\}:=\sigma\big(\,[\,a\,,\,b\,]\,\big)$. 
 This Schouten algebra $\text{gr}{\mathcal A}$ will be called the \textit{Poisson limit of the almost-commutative algebra}. 
The Schouten algebra $\mathcal{S}(M)$ of principal symbols is isomorphic to the Poisson limit of the almost-commutative algebra $\mathcal{D}(M)$ of differential operators: $\mathcal{S}(M)\cong\text{gr}\,\mathcal{D}(M)$.

\subsection{Automorphisms}\label{autom}

\paragraph{Finite automorphisms.} An automorphism $F:\mathcal{A}\stackrel{\sim}{\to}\mathcal{A}$ of an algebra $\mathcal{A}$ is an isomorphism of the vector space $\mathcal{A}$ into itself that preserves the product (\textit{i.e.} $F(a_1a_2)=F(a_1)F(a_2)$, $\forall a_1,a_2\in\mathcal{A}$). An automorphism of a filtered algebra $\mathcal A$ must also preserve the filtration (\textit{i.e.} $F({\mathcal{A}_i})\subset\mathcal{A}_i$, $\forall i\in\mathbb{N}$). An automorphism of a Poisson algebra $\mathcal{P}$ must preserve both products: the commutative product \textit{and} the Poisson bracket (\textit{i.e.} $F(f_1\cdot f_2)=F(f_1)\cdot F(f_2)$ and $F(\,\{f_1,f_2\}\,)=\{\,F(f_1),F(f_2)\,\}$ for any $f_1,f_2\in\mathcal{P}$). 

\paragraph{Diffeomorphisms.} A diffeomorphism of a smooth manifold $M$ is equivalent to an automorphism of the algebra $C^\infty(M)$ of smooth functions on $M$. In particular, a symplectomorphism of a symplectic manifold $\mathcal{M}$ can be defined as an automorphism of the Poisson algebra $C^\infty(\mathcal{M})$.

\paragraph{Infinitesimal automorphisms.} The derivations of an algebra are its infinitesimal automorphisms. An inner derivation of an associative (or Lie) algebra is an endomorphism $ad_a$ of the algebra defined in terms of the commutator (or Lie bracket) as $ad_a(b):=[a,b]$. The infinitesimal automorphisms of a filtered algebra are those derivations that preserve the filtration.

\paragraph{Lie derivatives along vector fields.} Let $\mathcal{L}_Xf$ denote the Lie derivative of the function $f\in C^\infty(M)$ along the vector field $X\in\mathfrak{X}(M)$. The vector fields $X\in\mathfrak{X}(M)$ on a manifold $M$ are in one-to-one correspondence with the derivations, $\mathcal{L}_X:f\mapsto \mathcal{L}_Xf$, of the algebra $C^\infty(M)$ of functions on $M$.  Let $\mathcal{L}_X Y=[X,Y]=ad_XY$ denote the Lie derivative of the vector field $Y\in\mathfrak{X}(M)$ along the vector field $X\in\mathfrak{X}(M)$. The vector fields $X\in\mathfrak{X}(M)$ on a manifold $M$ are in one-to-one correspondence with the inner derivations, $ad_X:Y\mapsto \mathcal{L}_XY$, of the Lie algebra $\mathfrak{X}(M)$ of vector fields on $M$. 
Symplectic vector fields on a symplectic manifold $\mathcal{M}$ are derivations of the Poisson algebra $C^\infty(\mathcal{M})$ for both its commutative product \textit{and} its Poisson bracket. In particular, Hamiltonian vector fields are inner derivations of the Poisson algebra $C^\infty(\mathcal{M})$. 

\paragraph{Higher-spin Lie derivatives along differential operators.} By analogy, the inner derivation $ad_{\hat{X}}:\hat{Y}\mapsto ad_{\hat{X}}\hat{Y}=[\hat{X},\hat{Y}]$ of the associative algebra ${\mathcal D}(M)$ of differential operators will be called the \textit{higher-spin Lie derivative along the differential operator} $\hat{X}\in{\mathcal D}(M)$. Note that if $\hat{X}\in{\mathcal D}^r(M)$ is a differential operator of order $r$, then the higher-spin Lie derivative $ad_{\hat X}:{\mathcal D}^q(M)\to{\mathcal D}^{q+r-1}(M)$ along $\hat{X}$ increases the order by $r-1$. Therefore, higher-spin Lie derivatives along first-order differential operators define infinitesimal automorphisms of the filtered algebra ${\mathcal D}(M)$ of differential operators, but
higher-order differential operators do \textit{not} define infinitesimal automorphisms of the \textit{filtered} algebra ${\mathcal D}(M)$ because they increase the order. This property is the root of the no-go theorems.

\paragraph{Obstruction to integrability.} A global flow on a manifold $M$ is nothing but an action of the additive group $\mathbb R$ on the manifold $M$. A vector field $X$ on $M$ is complete if it generates a global flow on $M$, \textit{i.e.} a group morphism from the one-dimensional Lie group $\mathbb R$ to the infinite-dimensional group of diffeomorphisms of $M$. More algebraically, a vector field $X$ on $M$ is complete if it generates a group morphism 
\be
\exp(\bullet{\mathcal L}_{X})\,:\,\mathbb{R}\to Aut\big(\,C^\infty(M)\,\big)\,:\,t\mapsto\exp(\,t\,{\mathcal L}_{X}\,)
\ee 
from the additive group $\mathbb R$ to the group of automorphisms of the commutative algebra $C^\infty(M)$. Via conjugation, this defines a group morphism 
\be\label{flows}
\exp(\,\bullet\,ad_{X}\,)\,:\,\mathbb{R}\to Inn\big(\,\mathfrak{X}(M)\,\big)\,:\,t\mapsto\exp(\,t\,ad_{X}\,)
\ee
from the additive group $\mathbb R$ to the group of inner automorphisms of the Lie algebra $\mathfrak{X}(M)$ of vector fields on $M$.
 
\paragraph{Obstruction to integrability.} The generalisation $\exp(\,\bullet\,ad_{\hat{X}})$ of flows \eqref{flows} for first-order differential $\hat{X}\in{\mathcal D}^1(M)$ works analogously, in the sense that it defines inner automorphisms $\exp(\,t\,ad_{\hat{X}})$ of the filtered algebra ${\mathcal D}(M)$ of differential operators (for sufficiently small parameter $t$).
Its naive extension to higher-order differential operators, $\hat{X}\in{\mathcal D}^r(M)$ with $r>1$, is tantalising but it fails to be well-defined.
The higher-spin Lie derivative $ad_{\hat X}$ along a higher-order differential operator is a well-defined derivation of $\mathcal{D}(M)$, but it does not preserve the filtration because it increases the order (by a finite amount). 
In fact, its power $ad^n_{\hat X}$ (appearing in the Taylor series of $\exp(\,t\,ad_{\hat{X}})$ at $t=0$) increase the order of differential operators on which it acts by $n(r-1)$, which becomes arbitrarily large when $n\to\infty$ and $r>1$.
This explains why the tentative exponentiation $\exp(\,t\,ad_{\hat{X}})$ has a wild action: it shifts the order by infinity!
Therefore, if $\hat{X}$ is a higher-order differential operator then $\exp(\,t\,ad_{\hat{X}})$ sends differential operators to objects that are outside ${\mathcal D}(M)$.\footnote{There would have been a way out if the higher-spin Lie derivative was locally nilpotent. However, it has  been shown (\textit{cf.} Section 3 of \cite{Grabowski:2002}) that $ad_{\hat X}$ is locally nilpotent (more precisely: for any zeroth-order differential operator $\hat{f}$, there exists a positive integer $n$ such that $ad^n_{\hat X}\hat{f}=0$) iff ${\hat X}$ is a differential operator of order zero. This is the crucial technical lemma behind
the no-go theorems.}

\vspace{3mm}

The above simple argument shows explicitly that, in order to define higher-spin diffeomorphisms, the space of differential operators should be completed, in order to include generalised differential operators with an infinite number of derivatives .

\section{Symplectomorphisms of the cotangent bundle}\label{symplectomcotgt}

The obstruction to the integrability of inner derivations of the associative algebra $\mathcal{D}(M)$ of differential operators to one-parameter group of inner automorphisms is not a ``quantum'' feature since the same applies for the Poisson algebra $\mathcal{S}(M)$ where  there is a similar obstruction to the integrability of inner derivations. In fact, the ``classical'' case shows a clear way out of the no-go theorem of Grabowski and Poncin.  The aim of this section is to use the cotangent bundle in order to explain geometrically the origin of the problem and its solution.

In order to demystify the subtleties at hand and gain some geometric intuition, they are illustrated on elementary examples in this section. On the way, basic concepts and results in symplectic geometry are reviewed. 

\subsection{Lagrangian submanifolds.}

Consider a submanifold $\mathcal{N}\subset\mathcal{M}$ of a symplectic manifold $\mathcal M$. It is defined by an embedding $i:\mathcal{N}\hookrightarrow\mathcal{M}$. A submanifold such that the pullback of the symplectic two-form $\Omega$ on $\mathcal M$ along this embedding vanishes identically on $\mathcal N$ (that is to say: $i^*\Omega=0$) is called an \textit{isotropic submanifold}. A maximal isotropic submanifold $\mathcal{N}\subset\mathcal{M}$ of a finite-dimensional symplectic manifold $\mathcal M$,
\be
i:\mathcal{N}\hookrightarrow\mathcal{M}\,,\quad i^*\Omega=0\,,\qquad 2\,\,\text{dim}\,\mathcal{N}\,=\,\text{dim}\,\mathcal{M}\,,
\ee
is called a \textit{Lagrangian submanifold}.

\vspace{3mm}

\noindent{\small\textbf{Example (Zero section of the cotangent bundle)\,:} 
The canonical fibration 
\be\label{canofibr}
\tau\,:\,T^*M\twoheadrightarrow M\,:\,(x^\mu,p_\nu)\mapsto x^\mu
\ee
of the cotangent bundle possesses a canonical section (as any vector bundle)\,: the zero section, \textit{i.e.} 
\be
\zeta\,:\,M\hookrightarrow T^*M\,:\,x^\mu\mapsto (x^\mu,0)\,.
\ee 
The pullbacks of the fibration and of the zero section define, respectively, the canonical embedding (of the subspace of functions on the base)
\be\label{canoT^*M}
\tau^*\,:\,C^\infty(M)\hookrightarrow C^\infty(T^*M)\,:\,f(x)\mapsto X(x,p)=f(x)
\ee
and the canonical projector (onto the space of functions on the base)
\be\label{0sectT^*M}
\zeta^*\,:\,C^\infty(T^*M)\twoheadrightarrow C^\infty(M)\,:\,X(x,p)\mapsto X_0(x)=X(x,0)\,.
\ee
The fibres $\tau^{-1}(m)=T_m^*M$ of the canonical fibration and the graph $\zeta(M)\subset T^*M$ of the zero section are Lagrangian submanifolds of the cotangent bundle. 
}

\vspace{3mm}

\noindent{\small\textbf{Example (Symplectic vector space)\,:} Consider a finite-dimensional symplectic vector space $W$ with symplectic two-form $\Omega$. A Lagrangian subspace $V\subset W$ is such that the quotient $W/V$ isomorphic to the dual space $V^*$ of the Lagrangian subspace: $W/V\cong V^*$. The isomorphism is provided by the symplectic two-form itself:
\be
\Omega\,:\,W/V\stackrel{\sim}{\to} V^*\,:\,[w]\mapsto\Omega(w,\bullet)\,,
\ee 
where $w\in W$ is a representative of the equivalence class $[w]\in W/V$ (\textit{i.e.} $w\sim w+v$ for any $v\in V$).
}

\vspace{3mm}

Any differential one-form $\alpha\in\Omega^1(M)$ is, by definition, equivalent to a section
$\bar{\alpha}:M\hookrightarrow T^*M$ of the canonical fibration \eqref{canofibr} of the cotangent bundle. If the differential one-form on $M$ reads as $\alpha=\alpha_\mu(x)\,dx^\mu$ in some local coordinates $x^\mu$ on $M$, then the section reads as $p_\mu=\alpha_\mu(x)$ in the corresponding local Darboux coordinates $(x^\mu,p_\nu)$ on $T^*M$. 
The pullback of the tautological one-form $\theta\in\Omega^1(T^*M)$ on the cotangent bundle along the section $\bar{\alpha}:M\hookrightarrow T^*M$ identifies with the original differential one-form, $\bar{\alpha}^*\theta=\alpha$ (as is obvious in components since $\theta=p_\mu dx^\mu$).
A submanifold $L\subset T^*M$ of the cotangent bundle $T^*M$ projects diffeomorphically on the base $M$ iff it is the graph of a differential one-form $\alpha$ on the base $M$, \textit{i.e.} $L=\bar{\alpha}(M)$. 

Consider an exact symplectic manifold $\mathcal{M}$ with Liouville one-form $\theta$, \textit{i.e.} $\Omega=d\theta$.
An \textit{exact submanifold} is a Lagrangian submanifold $\mathcal{N}\subset\mathcal{M}$ such that the pullback of the Liouville one-form by the embedding $i:\mathcal{N}\hookrightarrow\mathcal{M}$ is exact (\textit{i.e.} $i^*\theta=dH$ for a function $H$ on $\mathcal{N}$).

\vspace{3mm}
\noindent{\small\textbf{Example (Generating function)\,:} 
Consider a function $H\in C^\infty(M)$ on the manifold $M$. Its differential $\alpha=dH\in\Omega^1(M)$ defines a section of the cotangent bundle, which reads as $p_\mu=\partial_\mu H(x)$ in Darboux coordinates.
The graph $dH(M)\subset T^*M$ of such a section is an exact Lagrangian submanifold of the cotangent bundle. In such case, the function $H$ on the base manifold is called a \textit{generating function}.
}

\vspace{3mm}

Putting everything together, a submanifold $L\subset T^*M$ of the cotangent bundle $T^*M$ projecting diffeomorphically on the base manifold 
is Lagrangian (respectively, exact) iff it is the graph of a closed (respectively, exact) differential one-form $\alpha$ on the base $M$, \textit{i.e.} $L=\alpha(M)$ with $d\alpha=0$ (respectively, $\alpha=dH$). 

\subsection{Flows of Hamiltonians of degree zero}

\textit{The Hamiltonian flows of the cotangent bundle generated by Hamiltonians of degree zero in the momenta produce vertical symplectomorphisms mapping the zero section to exact submanifolds.}

\vspace{2mm}
\noindent{\small\textbf{Proof:} To see this, pick a point $p\in T_m^*M$ of the cotangent bundle, \textit{i.e.} a tangent covector $p$ at $m\in M$. Consider a differential one-form $A=A_\mu(x)\,dx^\mu\in\Omega^1(M)$.
The translations $p\mapsto p+A|_m$ by the differential one-form on $M$ read in fibre coordinates as
\be
x^\mu\mapsto x^\mu\,,\qquad p_\nu\mapsto p'_\nu=p_\nu+A_\nu(x)\,.
\ee
These translations are vertical diffeomorphisms of the cotangent bundle $T^*M$. Moreover, they are (exact) symplectomorphisms iff the differential one-form $A$ is closed (exact). In physical terms, they correspond to a minimal coupling to an electromagnetic field with vanishing fieldstrength (pure gauge).  
In geometrical terms, the corresponding vertical flow is symplectic/Hamiltonian iff it maps the zero section to Lagrangian/exact submanifolds. If this vertical flow is Hamiltonian, then the corresponding Hamiltonian $H(x)$ is a function on the base $M$ independent of the momenta. This ends the proof.} $\square$

\vspace{3mm}
\noindent{\small\textbf{Example (Vertical affine symplectomorphisms)\,:} Consider a vector space $V$ with basis $\{\texttt{e}_a\}$ and  Cartesian coordinates $y^a$. The coordinates on the cotangent bundle $T^*V\cong V\oplus V^*$ are $(y^a,p_b)$\,. 
The Hamiltonian flow of $T^*V$ generated by a homogenous Hamiltonian $H(y,p)=\alpha_a\,y^a$, of degree one in the positions $y$ and zero in the momenta $p$, produce vertical translations, $(y^a,p_b)\mapsto (y^a,p_b+t\,\alpha_b)$, mapping the zero section $\zeta(V)=V\oplus\,0$ (of equation $p_a=0$) to parallel affine subspaces (of equation $p_a=-t\,\alpha_a$). The Hamiltonian flow of $T^*V$ generated by the homogenous Hamiltonian $H(y,p)=\frac12\alpha_{ab}\,y^ay^b$, of degree two in the positions and zero in the momenta, produce linear vertical symplectomorphisms, $(y^a,p_b)\mapsto (y^a,p_b+t\,\alpha_{bc}\,y^c)$, mapping the zero section $p_a=0$ to linearly independent exact subspaces (of equation $p_a=-t\,\alpha_{ab}\,y^b$). The group of all vertical affine symplectomorphisms is isomorphic to the abelian group $V^*\oplus(V^*\odot V^*)$.
}

\subsection{Lagrangian foliations}

A foliation of a symplectic manifold $\mathcal M$ whose leaves are Lagrangian submanifolds is called a \textit{Lagrangian foliation} (in symplectic geometry) or a \textit{polarisation} (in geometric quantisation) of a symplectic manifold.

\vspace{3mm}
\noindent{\small\textbf{Example (Vertical polarisation of the cotangent bundle)\,:} 
The fibration $\tau:T^*M\twoheadrightarrow M$ defining the cotangent bundle provides an example of Lagrangian foliation (since each cotangent space $T^*_mM$ is a Lagrangian submanifold of the cotangent bundle $T^*M$), called the \textit{vertical polarisation of the cotangent bundle}.
Note that the cotangent bundle may admit other Lagrangian foliations than this canonical one.
}

\vspace{3mm}
\noindent{\small\textbf{Example (Symplectic vector space)\,:} The direct sum $V\oplus V^*$ of a finite-dimensional vector space $V$ and its dual $V^*$ is a finite-dimensional symplectic vector space endowed with a canonical symplectic two-form $\Omega$ defined by $\Omega(v\oplus\alpha,w\oplus\beta)=\alpha(w)-\beta(v)$ for all $v,w\in V$ and $\alpha,\beta\in V^*$. 
Conversely, any finite-dimensional symplectic vector space $W$ is isomorphic to a direct sum $V\oplus V^*$ of a Lagrangian subspace space $V\subset W$ with its dual $V^*$. 
In fact, any finite-dimensional symplectic vector space $W$ may be decomposed as 
the direct sum $V\oplus L$ of a Lagrangian subspace $V\subset W$ and another Lagrangian subspace $L\subset W$ complementary to $V$. Moreover, the latter subspace $L\cong W/V$ is isomorphic to the dual of the former subspace $V$, \textit{i.e.} $L\cong V^*$.
Any Lagrangian subspace $L\subset W$ of a symplectic vector space $W$ defines a Lagrangian foliation by all affine subspaces parallel to $L$.}

\vspace{3mm}
\noindent{\small\textbf{Example (Cotangent bundle of a vector space)\,:} 
The cotangent bundle of the vector space $V$ is a finite-dimensional symplectic vector space which decomposes as the direct sum $T^*V=V\oplus T^*_0V$ of the base space $V$ and the cotangent space $T^*_0V\cong V^*$ at the origin. 
Actually, any finite-dimensional symplectic vector space $W$ identifies with the cotangent bundle $T^*V$ of a finite-dimensional vector space $V$ upon a choice of polarisation $W=V\oplus L$. 
}

\subsection{Flows of Hamiltonians of degree one}

\textit{The Hamiltonian flows of the cotangent bundle generated by homogeneous Hamiltonians of degree one in the momenta produce symplectomorphisms obtained by lifting diffeomorphisms of the base. They preserve both the vertical polarisation and the zero section of the cotangent bundle.}

\vspace{2mm}
A symplectomorphism of the cotangent bundle $T^*M$ coincides with the lift of a diffeomorphism of the base manifold $M$ iff it preserves the tautological one-form $\theta$. In Darboux coordinates, this means that $p'_\mu dx'{}^{\mu}=p_\mu dx^{\mu}$ iff the change of coordinates takes the form 
\be
x^\mu\mapsto x'{}^{\mu}(x)\,,\qquad p_\nu\mapsto p'_\nu=\frac{\partial x^{\rho}}{\partial x'{}^{\nu}}\,p_\rho.
\ee
Consider a Hamiltonian flow of the cotangent bundle $T^*M$. The following statements are equivalent: the flow
\begin{itemize}
	\item[(a)] preserves the tautological one-form,
	\item[(b)] is the lift of a flow on the base manifold $M$ generated by a base vector field, $\hat{X}=X^\mu(x)\,\partial_\mu$,
	\item[(c)] is generated by a homogenous Hamiltonian of degree one in the momenta, $H(x,p)=X^\mu(x)\,p_\mu$.
\end{itemize}

\vspace{3mm}
\noindent{\small\textbf{Example (Lift of affine transformation)\,:} Consider a vector space $V$ with Cartesian coordinates $y^a$. Darboux coordinates on the cotangent bundle $T^*V=V\oplus V^*$ are $(y^a,p_b)$\,. The Hamiltonian flows of $T^*V$ generated by Hamiltonians with affine dependence in the positions and linear dependence in the momenta,
\be
H(y,p)=(\lambda^b\,+\,\lambda^b{}_a y^a)\,p_b\,,
\ee
are lifts 
\be
v\oplus\alpha\,\mapsto\,v'\oplus\alpha'\,=\,\big(\,\ell+\Lambda(v)\,\big)\,\oplus\,\big(\,(\Lambda^T)^{-1}(\alpha)\,\big)
\ee
of affine transformations $v\mapsto v'=\ell+\Lambda(v)$ of the base $V$, where $\ell\in V$ defines a translation and $\Lambda\in GL(V)$ defines a general linear transformation whose transpose is denoted $\Lambda^T\in GL(V^*)$.
}

\vspace{3mm}

More generally, the following statements are equivalent: a Hamiltonian flow of the cotangent bundle of a manifold $M$
\begin{itemize}
	\item[(a)] preserves the vertical polarisation, \textit{i.e.} it maps cotangent spaces $T^*_mM$ to cotangent spaces  $T^*_{m'}M$,
	\item[(b)] is the composition of a vertical symplectomorphism and the lift of a diffeomorphism of the base $M$,
	\item[(c)] is generated by a Hamiltonian of degree one in the momenta, $H(x,p)=X^\mu(x)\,p_\mu+f(x)$.
\end{itemize}

\vspace{3mm}
\noindent{\small\textbf{Example (Affine symplectomorphisms)\,:} Consider again the cotangent bundle $T^*V$ of a vector space $V$. Its affine symplectomorphisms span the affine symplectic group $ISp\,(V\oplus V^*)$.
The subgroup of \textit{affine} symplectomorphisms that preserve the vertical polarisation and the Lagrangian subspace $V\oplus 0\subset T^*V$ is isomorphic to the affine group $IGL(V)$.
Finally, the group of affine symplectomorphisms that preserve the vertical polarisation also contains an abelian subgroup: the subgroup $V^*\oplus(V^*\odot V^*)$ of vertical affine symplectomorphisms.
In fact, the group of affine symplectomorphisms preserving the vertical polarisation is isomorphic to the semidirect product 
$$
IGL(V)\ltimes \big(\,V^*\oplus(V^*\odot V^*)\,\big)\,.
$$ 
}

\subsection{Flows of Hamiltonians of higher degree}

\textit{The Hamiltonian flows of the cotangent bundle that do not preserve the vertical polarisation are
generated by homogeneous Hamiltonians of degree strictly greater than one in the momenta. The converse is also true.}

\vspace{2mm}
A symplectomorphism of the tangent bundle, $\Phi:T^*M\stackrel{\sim}{\to}T^*M$, that does not preserve the canonical fibration $\tau:T^*M\twoheadrightarrow M$ does, nevertheless, preserves the symplectic structure by definition. Thus, it maps the cotangent space $T_m^*M$ at a point $m$ to a Lagrangian submanifold $\Phi(T_m^*M)$. The latter can be taken as fibre over $\Phi(m)$. Thus, any symplectomorphism $\Phi$ defines a Lagrangian foliation (in general, distinct from the canonical one). 

\vspace{3mm}
\noindent{\small\textbf{Example (Linear changes of polarisation)\,:} Consider once again the cotangent bundle $T^*V$ of a vector space $V$.
The Hamiltonian flows generated by Hamiltonians, $H(y^a,p_b)=\frac12\,Y^{ab}p_ap_b$, independent of the positions and quadratic in the momenta are horizontal in the sense that they preserve the zero section $p_b=0$, in fact they read as $(y^a,p_b)\mapsto (y^a+t\,Y^{ac}p_c,p_b)$. 
The vertical polarisation of $T^*V$  by the affine subspaces $y^a=y^a_0$\,, parallel to the cotangent space $T^*_0V$ at the origin (of equation $y^a=0$), is mapped to the polarisation of $T^*V$ by affine subspaces of equation $y^a=y_0^a+t\,Y^{ab}p_b$\,.
The group of horizontal changes of polarisation is an abelian subgroup $V\odot V\subset Sp\,(V\oplus V^*)$ of the group of linear symplectomorphisms. The affine symplectomorphisms of the cotangent bundle $T^*V$ of the vector space $V$ are summarised in the table \ref{symplecto}.
}

\begin{table}
\begin{center}
\footnotesize
\begin{tabular}{
|c|c|c|c|c|}
\hline
Lie algebra & Dimension & Exact symplectomorphisms & Basis of Hamiltonian & Basis of\\
 &  &  & vector fields & Hamiltonians\\
\hline\hline
&&&&\\
$V^*$ & $n$ & Vertical translations & $\frac{\partial}{\partial p_a}$ & $x^a$ \\
&&&&\\
\hline
&&&&\\
$V^*\odot V^*$ & $\frac{n(n+1)}2$ & Linear vertical symplectom. & $x^b\,\frac{\partial}{\partial p_c}+x^c\,\frac{\partial}{\partial p_b}$ & $\frac12\,x^b\,x^c$ \\
&&&&\\
\hline
&&&&\\
$V^*\oplus(V^*\odot V^*)$ & $\frac{n(n+3)}2$ & Affine vertical symplectom. & $\frac{\partial}{\partial p_a}$, $x^b\frac{\partial}{\partial p_c}+x^c\,\frac{\partial}{\partial p_b}$ & $x^a$, $x^b\,x^c$ \\
&&&&\\
\hline\hline
&&&&\\
$V$ & $n$ & Horizontal translations & $\frac{\partial}{\partial x^a}$ & $p_a$ \\
&&&&\\
\hline
&&&&\\
$\mathfrak{gl}(V)$ & $n^2$ & Lift of linear transformations  & $x^b\,\frac{\partial}{\partial x^c}-p_c\,\frac{\partial}{\partial p_b}$ & $x^b p_c$ \\
&&&&\\
\hline
&&&&\\
$\mathfrak{igl}(V)$ & $n^2+n$ & Lift of affine transformations & $\frac{\partial}{\partial x^a}$, $x^b\frac{\partial}{\partial x^c}-p_c\frac{\partial}{\partial p_b}$ & $p_a$, $x^b p_c$ \\
&&&&\\
\hline\hline
&&&&\\
$V\odot V$ & $\frac{n(n+1)}2$ & Linear changes of polarisation & $p_a\frac{\partial}{\partial x^b}+p_b\frac{\partial}{\partial x^a}$ & $\frac12\,p_a\,p_b$ \\
&&&&\\
\hline\hline
&&&&\\
$\mathfrak{isp}(T^*V)$ & $2n^2+3n$ & Affine symplectomorphisms & Affine vector fields & degree 2  \\
&&&&\\
\hline
\end{tabular}
\end{center}
\caption{Affine symplectomorphisms of the tangent bundle $T^*V\cong V\oplus V^*$ of a vector space $V$ of dimension $n$}
\label{symplecto}
\end{table}

\vspace{3mm}
\noindent{\small\textbf{Example (Changes of polarisation)\,:} On the cotangent bundle $T^*M$ of a manifold $M$, one may also consider the linear changes of polarisations $(x^\mu,p_\nu)\mapsto (x^\mu+t\,Y^{\mu\rho}p_\rho,p_\nu)$ in some Darboux coordinates. One can explicitly see that such transformations do not preserve the space $\mathcal{S}(M)$ of symbols. For instance, the symbol $f(x^\mu,p_\nu)=\exp(k_\mu x^\mu)$ of degree zero in the momenta is mapped to the function $f(x^\mu+t\,Y^{\mu\rho}p_\rho,p_\nu)=\exp(k_\mu x^\mu)\exp(\,t\,Y^{\nu\rho}k_\nu p_\rho)$ which is manifestly \textit{not} polynomial in the momenta for $t\neq 0$.
}

\subsection{Summary}

Combining all those observations sheds some light on the Grabowski-Poncin theorem on the scarcity of finite automorphisms of the Poisson algebra of symbols with respect to its infinitesimal automorphisms. The one-parameter groups of inner automorphisms of the Poisson algebra $\mathcal{S}(M)$ of symbols on the manifold $M$ are necessarily generated by symbols of degree one. 
Retrospectively, this is somewhat natural since only symbols of degree one generate Hamiltonian symplectomorphisms of $T^*M$ preserving the vertical polarisation, the latter being instrumental in the intrinsic definition of the algebra $\mathcal{S}(M)$ of symbols. Breaking the vertical polarisation simultaneously destroys the polynomiality in momenta.
Nevertheless, these Hamiltonian vector fields on $T^*M$ are symplectomorphisms, therefore they will map the vertical polarisation to another choice of polarisation, with respect to which the pullback of symbols will be polynomial in the new ``momenta''.

\section{Quantisation of the cotangent bundle: Differential operators as symbols and vice versa}\label{quantisations}

The Poisson algebra $C^\infty(T^*M)$ offers a suitable completion\footnote{The Stone–Weierstrass theorem ensures that $\mathcal{S}(M)$ is dense inside $C^\infty(T^*M)$.} of the Schouten algebra $\mathcal{S}(M)$ of symbols on which symplectic diffeomorphisms of the cotangent bundle admit an algebraic definition as automorphisms of the Poisson algebra.
In order to construct a similar completion of the almost-commutative algebra $\mathcal{D}(M)$ of differential operators on should first describe it as a quantisation of its Poisson limit $\mathcal{S}(M)$.

\subsection{Quantisation of the cotangent bundle}\label{compatcond}

\paragraph{Quantisation of Schouten algebras.} 
An isomorphism $q:\mathcal{S}\stackrel{\sim}{\to}\mathcal{A}$ of vector spaces from a Schouten algebra $\mathcal{S}$ to an almost-commutative algebra $\mathcal{A}$ such that its restrictions $q_i:\mathcal{S}_i\hookrightarrow\mathcal{A}_i$ composed with the ones of the principal symbol map $\sigma_i:{\mathcal A}_i\twoheadrightarrow\text{gr}_i{\mathcal A}$ define
\begin{enumerate}
	\item[(i)] a collection of isomorphisms $\widetilde{q}_i=\sigma_i\circ q_i:\mathcal{S}_i\stackrel{\sim}{\to}\text{gr}_i\mathcal{A}$ of vector spaces, and
	\item[(ii)] an isomorphism $\widetilde{q}\,:\,\mathcal{S}\stackrel{\sim}{\to}\text{gr}\,\mathcal{A}$ of Schouten algebras between $\mathcal{S}$ and the Poisson limit of $\mathcal{A}$,
\end{enumerate}
will be called a \textit{quantisation of the Schouten algebra} $\mathcal{S}$ \textit{into the almost-commutative algebra} $\mathcal{A}$.\footnote{If the quantised Schouten algebra happens to be equal to the Poisson limit of the almost-commutative algebra (\textit{i.e.} $\mathcal{S}=\text{gr}{\mathcal A}$) then one also requires that the restriction $q_i$ of the quantisation is a section of the restriction $\sigma_i$ of the principal symbol map ($\sigma_i\circ q_i=id_{\mathcal{S}_i}$).}
 
\paragraph{Example (Universal enveloping algebra of a Lie algebra).} Let $\mathfrak{g}$ be a Lie algebra. The universal enveloping algebra $U(\mathfrak{g})$ is almost-commutative. The Poincar\'e-Birkhoff-Witt map
\be\label{pbw}
pbw\,:\,\odot(\mathfrak{g})\stackrel{\sim}{\to} U(\mathfrak{g})\,:\,y_1\odot\cdots\odot y_n\mapsto \frac1{n!}\sum\limits_{s\in\mathfrak{S}_n}y_{s(1)}\cdots y_{s(n)}
\ee
is a quantisation of the symmetric algebra $\odot(\mathfrak{g})$ of the Lie algebra.

\paragraph{Quantisation of the cotangent bundle.} A quantisation $Q:\mathcal{S}(M)\stackrel{\sim}{\to}\mathcal{D}(M)$ from the Schouten algebra of symbols on $M$ into the almost-commutative algebra of differential operators on $M$ will be called a \textit{quantisation of the cotangent bundle} $T^*M$.

\paragraph{Transfer of structures.}
Note that a quantisation $q:\mathcal{S}\stackrel{\sim}{\to}\mathcal{A}$ allows to transfer each structure on the other: On the one hand, $\mathcal{A}$ inherits the grading of $\mathcal S$ as follows:
$\mathcal{A}|_i=q(\mathcal{S}_i)$\,.
On the other hand, one may induce an associative product $\star$ on $\mathcal{S}$ from the associative product $\circ$ of $\mathcal A$, as follows:
$q(f\star g)=q(f)\circ q(g)$ for all $f,g\in\mathcal S$\,.
In this way, the vector space $\mathcal{S}$ becomes endowed with a structure of almost-commutative algebra. One may decompose the induced product $\star$ via the grading of $\mathcal{S}$ as
\be\label{hbar1}
\star=\sum\limits_{k=0}^\infty\,\star_k\,,\qquad 
\star_k\,:\,\,\mathcal{S}_i\otimes\mathcal{S}_j\to\mathcal{S}_{i+j-k}\,.
\ee
where $\star_k$ decreases the grading by $k$.
Let $\cdot$ and $\{\,,\,\}$ denote respectively the commutative product and the Lie bracket of the Poisson algebra $\mathcal S$. The condition (ii) implies that $\star_0$ is equal to the original commutative product and that the commutator bracket of $\star_1$ is equal to the original Poisson bracket, 
\be
\star_{{}_0}\,=\,\cdot\qquad\text{and}\qquad
[\,\,\,,\,\,]_{\star_1}\,=\,\{\,\,,\,\}\,\,.
\ee

\paragraph{Example (Symmetric algebra of a Lie algebra).} Let $\mathfrak{g}$ be a Lie algebra. The quantisation  
\eqref{pbw} induces a Poisson structure on its symmetric algebra $\odot(\mathfrak{g})$ for which $\star_{{}_0}=\odot$ and the Poisson bracket arises from the Lie bracket of $\mathfrak{g}$. Moreover, the explicit form of the induced product $\star$ is known \cite{Berezin}.

\subsection{Compatibility condition}

\paragraph{Rings over an associative algebra.} Let $\mathcal A$ and $\mathcal B$ be two associative algebras with respective units $1_{\mathcal A}$ and $1_{\mathcal B}$. An injective morphism $i:{\mathcal A}\hookrightarrow{\mathcal B}$ of associative algebras
will be called a \textit{unit map}.\footnote{Note that the injectivity can be assumed without loss of generality, in the sense that one can always focus on the quotient algebra ${\mathcal A}/\ker\,i$. The terminology originates from the fact that, in particular, the unit map relates the two units in the sense that $i(1_{\mathcal A})=1_{\mathcal B}$).} An associative algebra $\mathcal B$ endowed with a unit map $i:{\mathcal A}\hookrightarrow{\mathcal B}$ is called a ring $\mathcal B$ over the base algebra $\mathcal A$ (or $\mathcal A$-ring for short). Equivalently, $\mathcal B$ admits a subalgebra isomorphic to $\mathcal A$\,: the image $i({\mathcal A})\subseteq\mathcal B$.
If the image of the unit map belongs to the centre, $i({\mathcal A})\subseteq Z({\mathcal B})$, then $\mathcal B$ is called an algebra over $\mathcal A$ (or $\mathcal A$-algebra).

\paragraph{Functions vs Differential operators of order zero.} The algebra $\mathcal{D}(M)$ of differential operators is a $C^\infty(M)$-ring. 
In fact, any function $f\in C^\infty(M)$ defines a differential operator of order zero, $\hat{f}\in\mathcal{D}^0(M)$, acting on $ C^\infty(M)$ by multiplication by $f$\,, \textit{i.e.} $\hat{f}:g\mapsto f\cdot g$. 
This provides a canonical unit map
\be\label{hatbull}
\hat{\bullet}\,:\, C^\infty(M)\hookrightarrow\mathcal{D}(M)\,:\,f\mapsto\hat{f}\,.
\ee
Note that $\hat{1}_{ C^\infty(M)}=id_{ C^\infty(M)}\,$.

\paragraph{Characters of rings over an algebra.} Consider an $\mathcal A$-linear map $\pi:{\mathcal B}\twoheadrightarrow{\mathcal A}$ which is a retraction of the unit map $i:{\mathcal A}\hookrightarrow{\mathcal B}$ and which relate the unit elements, \textit{i.e.}
\be\label{char1}
\pi\big(i(a)\star b\big)=a\cdot\pi(b)\,, \quad \pi \circ i=id_{\mathcal A}\,, \quad\pi(1_{\mathcal B})=1_{\mathcal A}\,.
\ee 
If this map is such that 
\be\label{char2}
\pi(b_1\star b_2)\,=\,\pi\Big(b_1\star i\big(\pi(b_2)\big)\Big)\,,\qquad\forall b_1,b_2\in{\mathcal B}\,,
\ee
then it is called a \textit{character on the} $\mathcal A$-\textit{ring} $\mathcal B$.
A character such that
\be\label{nondegchar1}
b_1\in{\mathcal B}\,:\quad\pi(b_1\star b_2)=0\,,\quad\forall b_2\in{\mathcal B}\qquad\Longleftrightarrow\qquad b_1=0\,,
\ee
or, equivalently\footnote{The equivalence between \eqref{nondegchar1} and \eqref{nondegchar2} holds because of \eqref{char2} and the surjectivity of $\pi:{\mathcal B}\twoheadrightarrow{\mathcal A}$.}, such that
\be\label{nondegchar2}
b\in{\mathcal B}\,:\quad\pi\big(\,b\star i(a)\,\big)=0\,,\quad a\in{\mathcal A}\qquad\Longleftrightarrow\qquad b=0\,,
\ee
will be called a \textit{non-degenerate character on the} $\mathcal A$-\textit{ring} $\mathcal B$.

\paragraph{Example (Cotangent bundle).} The pullback $\tau^*$ of the fibration $\tau:T^*M\twoheadrightarrow M$ of the cotangent bundle defines a unit map \eqref{canoT^*M} endowing $C^\infty(T^*M)$ with a structure of $C^\infty(M)$-ring.
The pullback $\zeta^*$ of the zero section $\zeta:M\hookrightarrow T^*M$ defines a degenerate character on the $C^\infty(M)$-ring $C^\infty(T^*M)$. The same applies for the (co)restriction of these maps to the subalgebra ${\mathcal S}(M)\subset C^\infty(T^*M)$ of symbols, hence ${\mathcal S}(M)$ is also a $C^\infty(M)$-ring endowed with a degenerate character. 

\paragraph{Example (Differential operators of order zero vs Functions).} The action of differential operators on the constant function $1\in{\mathcal C}^\infty(M)$ defines a non-degenerate character on the $C^\infty(M)$-ring ${\mathcal D}(M)$ of differential operators,
\be\label{DM1}
\bullet[1]\,:\,{\mathcal D}(M)\twoheadrightarrow{\mathcal C}^\infty(M)\,:\,\hat{X}\mapsto X_0=\hat{X}[1]\,.
\ee
The function $X_0\in{\mathcal C}^\infty(M)$ defines a differential operator $\hat{X}_0\in{\mathcal D}^0(M)$ of order zero, which is the component of order zero of the differential operator $\hat{X}\in{\mathcal D}(M)$.
The composition $\hat{\zeta}^*$ of the character \eqref{DM1} followed by the unit map \eqref{hatbull} defines a surjective linear map
\be\label{zetatbullet}
\hat{\zeta}^*\,:\,{\mathcal D}(M)\twoheadrightarrow {\mathcal D}^0(M)\,:\,\hat{X}\mapsto \hat{X}_0
\ee
from the algebra ${\mathcal D}(M)$ of all differential operators onto the subalgebra ${\mathcal D}^0(M)$ of differential operators  of order zero.

\paragraph{Anchors of rings over an algebra.}
Consider a ring ${\mathcal B}$ over an algebra ${\mathcal A}$\,.
A morphism from the $\mathcal A$-ring ${\mathcal B}$ to the $\mathcal A$-ring $End({\mathcal A})$ of endomorphisms of ${\mathcal A}$,
\be\label{anchorB}
\hat{\bullet}\,:\,{\mathcal B}\to End({\mathcal A})\,:\,b\mapsto \hat{b}\,.
\ee
will be called an \textit{anchor of the} $\mathcal A$-\textit{ring} $\mathcal B$.\footnote{This terminology is borrowed from \cite{Xu:1999eh} where the anchor is defined for bialgebroids (an extra compatibility condition with the coproduct is added).} 
In other words,
\begin{itemize}
	\item[(i)] it is an algebra morphism, \textit{i.e.} it relates the product $\star$ in ${\mathcal B}$ to the product $\circ$ in $End({\mathcal A})$ one has:
\be\label{anchor1}
\widehat{b_1\star b_2}\,=\,\hat{b}_1\circ\hat{b}_2\,,\qquad\forall b_1,b_2\in{\mathcal B}\,,
\ee
In particular, an anchor \eqref{anchorB} defines a representation of the $\mathcal A$-ring ${\mathcal B}$ on its base algebra ${\mathcal A}$\,.
	\item[(ii)] it relates their unit maps, \textit{i.e.} the anchor extends the canonical isomorphism \eqref{hatbull} in the sense that 
\be\label{anchor3}
\widehat{i(a)}=\hat{a}\,,\qquad\forall a\in{\mathcal A}\,,
\ee
where ``hat'' stands, in the left-hand-side, for the anchor \eqref{anchorB} and, in the right-hand-side, for the canonical isomorphism \eqref{hatbull}. In particular, an anchor relates the unit elements, \textit{i.e.} 
\be\label{anchor2}
\hat{1}_{\mathcal B}=id_{\mathcal A}\,.
\ee 
\end{itemize}

\paragraph{Example (Cotangent bundle).} Given a quantisation of the cotangent bundle $T^*M$, the following square is commutative:
\be\label{extracondquantcom}
\begin{array}
[c]{ccc}%
\mathcal{S}(M)&\stackrel{Q}{\longrightarrow}&\mathcal{D}(M)\\
&&\\
\tau^*\big\uparrow&&\big\uparrow i\\
&&\\
 C^\infty(M)&\stackrel{\hat{\bullet}}{\longrightarrow}&\mathcal{D}^0(M)
\end{array}
\ee
where the vertical arrows are, respectively, the unit map \eqref{canoT^*M} of the $\mathcal{C}^\infty(M)$-ring $\mathcal{S}(M)$ and the 
embedding $i:\mathcal{D}^0(M)\hookrightarrow\mathcal{D}(M)$.
Therefore, any quantisation $Q:\mathcal{S}(M)\stackrel{\sim}{\to}\mathcal{D}(M)$ of the cotangent bundle is an injective anchor of the $\mathcal{C}^\infty(M)$-ring $\mathcal{S}(M)$ of symbols, whose image is the $\mathcal{C}^\infty(M)$-ring $\mathcal{D}(M)$ of differential operators.

\paragraph{Character $\equiv$ Anchor\,.}
The notions of anchor and character on $\mathcal A$-rings are actually equivalent to each other. On the one hand, from an anchor $\hat{\bullet}:{\mathcal B}\to End({\mathcal A})$ one may define a character $\pi:{\mathcal B}\twoheadrightarrow{\mathcal A}$ via
\be\label{pidef}
\pi(b)\,:=\,\hat{b}\,\big[1_{\mathcal B}\big]\,.
\ee
On the other hand, from a character $\pi:{\mathcal B}\twoheadrightarrow{\mathcal A}$ one may define an anchor $\hat{\bullet}:{\mathcal B}\to End({\mathcal A})$ via
\be
\hat{b}\,[a]\,:=\,\pi\big(b\star i(a)\big)\,.
\ee
One can check by a direct computation that the properties \eqref{char1}-\eqref{char2} of a character and the properties \eqref{anchor1}-\eqref{anchor3} of a character imply each other.
Note that the anchor is injective iff the character is non-degenerate.

\paragraph{Compatibility condition.} A quantisation of the cotangent bundle $T^*M$ defines an injective anchor. However, the corresponding character $\pi$ defined by \eqref{pidef} and the canonical character $\zeta^*$ defined in \eqref{0sectT^*M} on the $\mathcal{C}^\infty(M)$-ring $\mathcal{S}(M)$ may differ in general.
Consider the square:
\be\label{extracondquant}
\begin{array}
[c]{ccc}%
\mathcal{S}(M)&\stackrel{Q}{\longrightarrow}&\mathcal{D}(M)\\
&&\\
\zeta^*\big\downarrow&&\big\downarrow\hat{\zeta}^*\\
&&\\
 C^\infty(M)&\stackrel{\hat{\bullet}}{\longrightarrow}&\mathcal{D}^0(M)
\end{array}
\ee
where the vertical arrows are the canonical characters \eqref{0sectT^*M} and \eqref{zetatbullet} on the $\mathcal{C}^\infty(M)$-ring $\mathcal{S}(M)$ and the $\mathcal{D}^0(M)$-ring $\mathcal{D}(M)$, respectively. A quantisation of the cotangent bundle $T^*M$ such that the diagram \eqref{extracondquant} is commutative, will be called a \textit{compatible quantisation of the cotangent bundle} (in the sense that it is compatible with the canonical characters on those rings). 

\subsection{Examples of quantisation of the cotangent bundle}

Consider the short exact sequence of $C^\infty(M)$-modules
\be\label{sesdiffsym}
0\to \mathcal{D}^{k-1}(M)\stackrel{i_k}{\hookrightarrow}\mathcal{D}^{k}(M)\stackrel{\sigma_k}{\twoheadrightarrow}\odot^k\mathcal{T}(M)\to 0\,,
\ee
where the maps $i_k:\mathcal{D}^{k-1}(M)\hookrightarrow\mathcal{D}^{k}(M)$
define the filtration, and the maps $\sigma_k:\mathcal{D}^k(M)\twoheadrightarrow\odot^k\mathcal{T}(M)$ \eqref{prsymb} send a differential operator onto its principal symbol.
The short exact sequence \eqref{sesdiffsym} reflects the fact that the Poisson limit $\text{gr}\,\mathcal{D}(M)$ of the almost-commutative algebra $\mathcal{D}(M)$ of differential operators is isomorphic to the Schouten $\odot\mathcal{T}(M)$ of symmetric multivector fields. 

A linear splitting (\textit{i.e.} a splitting of vector spaces)
\be
0\leftarrow \mathcal{D}^{k-1}(M)\stackrel{r_k}{\twoheadleftarrow}\mathcal{D}^{k}(M)\stackrel{q_k}{\hookleftarrow}\odot^k\mathcal{T}(M)\leftarrow 0\,,\label{splitsesdiffsym}
\ee
of the short exact sequence \eqref{sesdiffsym} is equivalent to a quantisation 
\be\label{quantisationDM}
q\,:\,\odot\mathcal{T}(M)\stackrel{\sim}{\to}\mathcal{D}(M)\,:\,X\mapsto\hat{X}
\ee
of the Schouten algebra $\odot\mathcal{T}(M)$ of symmetric multivector fields into the almost-commutative algebra $\mathcal{D}(M)$ of differential operators.
More explicitly, for $k>0$ the sections $q_k$ of the principal symbol \eqref{prsymb} take the form
\be\label{qk}
q_k\,:\,\odot^k\mathcal{T}(M)\hookrightarrow\mathcal{D}^{k}(M)\,:\,X_k^{\mu_1\cdots \mu_k}(x)\,\partial_{\mu_1}\odot\,\cdots\,\odot\partial_{\mu_k}\mapsto\sum\limits_{r=0}^{k} Z_{X_k}^{\nu_1\cdots\nu_r}(x)\partial_{\nu_1}\cdots \partial_{\nu_r}
\ee
where the differential operator of order $k>0$ reads
\be\label{Zk}
\sum\limits_{r=0}^{k} Z_{X_k}^{\nu_1\cdots \nu_r}(x)\,\partial_{\nu_1}\cdots \partial_{\nu_r}=X^{\mu_1\cdots \mu_k}(x)\,\partial_{\mu_1}\cdots \partial_{\mu_k}+\sum\limits_{r=0}^{k-1} Z_{X_k}^{\nu_1\cdots \nu_r}(x)\,\partial_{\nu_1}\cdots \partial_{\nu_r}\,,
\ee 
with the coefficients $Z_{X_k}$ being linear in the components $X_k^{\mu_1\cdots \mu_k}(x)$.
Each such quantisation \eqref{quantisationDM} is compatible with the principal symbol, in the sense that $\sigma\circ q=id_{\odot\mathcal{T}(M)}$ (since $\sigma_k\circ q_k=id_{\odot^k\mathcal{T}(M)}$ by definition of a splitting). 
For a quantisation \eqref{quantisationDM} compatible with the canonical characters, the sums in \eqref{Zk} should start from $r=1$ for any $k>0$.

Obviously, for $k=0$ the short exact sequences \eqref{sesdiffsym} and \eqref{splitsesdiffsym} are degenerate and take the form
\be\label{sesdiffsym0}
0\to\mathcal{D}^0(M)\stackrel{\sim}{\rightarrow} C^\infty(M)\to 0\quad\text{and}\quad
0\leftarrow \mathcal{D}^0(M)\stackrel{\sim}{\leftarrow} C^\infty(M)\leftarrow 0\,.
\ee
In particular, $q_0=\hat{\bullet}:C^\infty(M)\stackrel{\sim}{\rightarrow}\mathcal{D}^0(M)$ is the canonical isomorphism
\eqref{hatbull} sending a function $f$ to the zeroth-order differential operator $\hat{f}$.
Similarly, the linear splitting \eqref{splitsesdiffsym} for $k=1$ is canonical
\be
0\leftarrow \mathcal{D}^{0}(M)\stackrel{r_1}{\twoheadleftarrow}\mathcal{D}^1(M)\stackrel{q_1}{\hookleftarrow}\mathcal{T}(M)\leftarrow 0\,,
\ee
where $q_1$ reinterprets vector fields as differential operators of order one.

The quantisation \eqref{quantisationDM} of the Schouten algebra $\mathcal{S}(M)$ defines a quantisation of the cotangent bundle $T^*M$
\be\label{quantisationDMap}
Q\,:\,\mathcal{S}(M)\stackrel{\sim}{\to}\mathcal{D}(M)\,:\,X\mapsto\hat{X}
\ee
which maps symbols of degree $k$ to differential operators of order $k$
\be\label{Qk}
Q_k\,:\,\mathcal{S}^k(M){\hookrightarrow}\mathcal{D}^{k}(M)\,:\,\sum\limits_{r=0}^k X_r^{\mu(r)}(x)\,p_{\mu_1}\cdots p_{\mu_r}\mapsto
\sum\limits_{0\leqslant s\leqslant r\leqslant k} Z_{X_r}^{\nu(s)}(x)\,\partial_{\nu_1}\cdots \partial_{\nu_s}\,,
\ee 
where the multi-index notation $\mu(r)\equiv\mu_1\cdots\mu_r$ was used for symmetric indices.
The inverse $\Sigma=Q^{-1}$ of a quantisation map \eqref{quantisationDMap}
will be called a \textit{symbol map},
\be
\Sigma\,:\,\mathcal{D}(M)\stackrel{\sim}{\to}\mathcal{S}(M)\,:\,\hat{X}\mapsto X\,.
\ee

A quantisation of the cotangent bundle such that each section $q_k$ is a differential operator will be called a \textit{differential quantisation}, \textit{i.e.} the $Z^{\nu(\ell)}_{X_k}$ are differential operators acting on the components of the principal symbols $X_k$, \textit{i.e.}
\be\label{sumZ}
Z^{\nu(\ell)}_{X_k}=\sum\limits_{m=0}^{\kappa(k,\ell)} Y^{\nu(\ell)\,|\,\rho(m)}_{\mu(k)}(x)\,\partial_{\rho_1}\cdots \partial_{\rho_m} X_k^{\mu(k)}\,,
\ee
where $\ell\leqslant k$ and the order $\kappa(k,\ell)$ of the differential operator depends on $k$ and $\ell$.
For quantisations corresponding to ``choices of ordering'' in some coordinate system $x^\mu$, the sections $q_k$ are differential operators of order $k$. Differential quantisations will be assumed to satisfy this extra condition. 
A differential quantisation $q:\odot\mathcal{T}(M)\stackrel{\sim}{\to}\mathcal{D}(M)$ which is $C^\infty(M)$-linear will be called a \textit{quantisation of normal type} (\textit{i.e.} all differential operators \eqref{sumZ} are of order zero, hence only the term $m=0$ is present in the sum). 

\vspace{3mm}
\noindent{\small\textbf{Example (Normal quantisation)\,:} Consider the manifold $M$ to be topologically trivial. Pick a global coordinate system $x^\mu$.\footnote{This quantisation is not canonical since, by construction, it depends explicitly on a choice of specific coordinate system. Nevertheless, this normal-type quantisation can be made geometrical (\textit{i.e.} globally well-defined and coordinate-independent) for a generic manifold $M$ by considering the following data: an affine connection on the base manifold $M$ (\textit{cf.} \cite{Bordemann:1997ep}). Retrospectively, the corresponding normal coordinates provide a privileged coordinate system.} Then, an example of compatible and normal-type quantisation is provided by the $C^\infty(M)$-linear maps
\be
q_k\,:\,X^{\mu_1\cdots \mu_r}(x)\,\partial_{\mu_1}\odot\cdots\odot\partial_{\mu_r}\mapsto X^{\mu_1\cdots \mu_r}(x)\,\partial_{\mu_1}\cdots \partial_{\mu_r}\,.
\ee
The corresponding quantisation of the cotangent bundle is the \textit{normal quantisation} sending symbols to the corresponding normal-ordered operators\,,
\be\label{standardquantisation}
Q_N\,:\,\sum\limits_{r=0}^k \frac1{r!}\,X^{\mu_1\cdots \mu_r}(x)\,p_{\mu_1}\cdots p_{\mu_r}\,\mapsto\,\sum\limits_{r=0}^k \frac1{r!}\,X^{\mu_1\cdots \mu_r}(x)\,\partial_{\mu_1}\cdots \partial_{\mu_r}\,.
\ee
The inverse map 
\be\label{standardsymbmap}
\Sigma_N\,:\,\hat{X}\mapsto X(x,p)\,=\,\exp(-p_\mu x^\mu)\,\hat{X}[\,\exp(p_\mu x^\mu)\,]
\ee
is the \textit{normal symbol map}.
}

\vspace{3mm}
\noindent{\small\textbf{Example (Weyl quantisation)\,:} The other paradigmatic example of differential quantisation of the cotangent bundle of a topologically trivial manifold is Weyl quantisation \cite{Weyl:1927}. It is based instead on the Weyl (\textit{i.e.} symmetric) ordering, instead of the normal ordering (in some Darboux coordinate system). The corresponding quantisation map is called the \textit{Weyl map} sending symbols to the corresponding Weyl-ordered operators. Its inverse is the symbol map called the \textit{Wigner map}.
For a review and explicit formulae, see \textit{e.g.} \cite{Bekaert:2009ud}. Note that the Weyl quantisation is not compatible with the canonical characters. For instance, the Weyl map sends the Weyl symbol $X(x,p)=x^\mu p_\mu$ to the Weyl-ordered operator $\hat{X}=\frac12(\hat{x}^\mu\circ\partial_\mu+\partial_\mu\circ\hat{x}^\mu)=x^\mu\partial_\mu+\frac12$ whose action on the unit gives $\hat{X}[1]=\frac12\neq 0=X(x,p=0)$.
}

\section{Quantisation of the cotangent bundle: Going beyond differential operators}\label{beyond}

\subsection{Quasi-differential operators}

A quantisation \eqref{quantisationDMap} of the cotangent bundle allows to endow the commutative algebra $\mathcal{S}(M)$ of symbols with a non-commutative product $\star$ inherited from the composition product $\circ$ of the almost-commutative algebra $\mathcal{D}(M)$ of differential operators. 

\paragraph{Strict product.} Let us assume that there exists an associative product $\star$ on the whole space $C^\infty(T^*M)$ of functions on the cotangent bundle $T^*M$ such that (i) it reduces to the above-mentioned product on the subspace $\mathcal{S}(M)\subset C^\infty(T^*M)$ of symbols on $M$ and (ii) the constant function $1\in C^\infty(T^*M)$ is the unit element for this product.
Such an associative product will be called a \textit{strict product} on $C^\infty(T^*M)$.
The space $C^\infty(T^*M)$ of smooth functions on the cotangent bundle endowed with a strict product $\star$ will be denoted $C_\star^\infty(T^*M)$. 
The canonical embedding \eqref{canoT^*M} of the commutative algebra $C^\infty(M)$ of functions on the base manifold inside the algebra $C^\infty(T^*M)$ of functions on the cotangent bundle $T^*M$ is a unit map of $C_\star^\infty(T^*M)$. 

\paragraph{Strict quantisation of the cotangent bundle.} Let us further assume that the $C^\infty(M)$-ring $C_\star^\infty(T^*M)$ is endowed with an injective anchor extending the quantisation map \eqref{quantisationDMap}. This hypothetical situation will  loosely\footnote{Usually, the term ``strict quantisation'' refers to one of the (many) mathematical approaches to the problem of quantisation and often refers to the axiomatisation by Rieffel (see \textit{e.g.} his book \cite{Rieffel}). Here, the term is understood in a non-technical sense.} be referred to as a \textit{strict quantisation of the cotangent bundle}.
In this ideal case, the $C^\infty(M)$-ring $C_\star^\infty(T^*M)$ could be interpreted as defining a completion of the almost-commutative algebra $\mathcal{D}(M)$ of differential operators $M$.

\paragraph{Quasi-differential operators.} The image of the  injective anchor will be denoted $\mathcal{Q}\mathcal{D}(M)$ and called the \textit{associative algebra of quasi-differential operators on the manifold} $M$. Let us motivate the terminology ``quasi-differential operator''\,: The term ``operator'' is justified by the fact that, by construction, the image $\mathcal{Q}\mathcal{D}(M)\subset End\big(\, C^\infty(M)\,\big)$ is spanned by linear operators on $C^\infty(M)$. More precisely, there is an isomorphism of associative algebras
\be\label{Qmap}
\hat{\bullet}\,:\,C_\star^\infty(T^*M)\stackrel{\sim}{\to}\mathcal{Q}\mathcal{D}(M)\,:\,X\mapsto\hat{X}
\ee
sending functions $X(x,p)$ on the cotangent bundle $T^*M$ on linear operators $\hat{X}$ acting on functions $f(x)$ on $M$. The map \eqref{Qmap} will be called a \textit{strict quantisation map} because it provides, by definition, an extension of some quantisation map $Q:\mathcal{S}(M)\stackrel{\sim}{\to}\mathcal{D}(M)$ sending symbols to differential operators.
In particular, the isomorphism \eqref{Qmap} extends the unit map \eqref{hatbull} sending functions on the base manifold $M$ to zeroth-order differential operators on $M$. 
The adjective ``differential'' to designate these operators comes from the fact that the vertical coordinates $p_\mu$ of generic functions $f(x,p)$ on the cotangent bundle $T^*M$ can loosely be interpreted as standing for partial derivatives $\partial_\mu$ while the adjective ``quasi'' underlines that the dependence is not polynomial in general. The table \ref{classvsquantalg} provides a comparison between the classical and quantum algebras of functions on the cotangent bundle that have been introduced so far. 

\begin{table}
\begin{center}
\begin{tabular}{
|c|c|c|}
\hline
& Classical & Quantum \\
\hline\hline
Algebra & Poisson algebra (symplectic) & Associative algebra (central) \\
 & $ C^\infty(T^*M)$ &  $\mathcal{Q}\mathcal{D}(M)$ \\
\hline
Elements & Functions on the cotangent bundle & Quasi-differential operators \\
 & $X(x,p)$ & $\hat{X}(x,\partial)$ \\
\hline\hline
Graded/Filtered & Schouten algebra & Almost-commutative algebra \\
subalgebra & $\mathcal{S}(M)$ &  $\mathcal{D}(M)$ \\
\hline
Elements & Symbols & Differential operators \\
 & $X(x,p)=\sum\limits_{r=0}^k \frac1{r!}\,X^{\mu_1\cdots \mu_r}(x)\,p_{\mu_1}\cdots p_{\mu_r}$ & $\hat{X}=\sum\limits_{r=0}^k \frac1{r!}\,X^{\mu_1\cdots \mu_r}(x)\,\partial_{\mu_1}\cdots \partial_{\mu_r}$ \\
\hline\hline
Commutative & Base algebra & Order zero subalgebra \\
subalgebra & $ C^\infty(M)\subset\mathcal{S}(M)$ &  $\mathcal{D}^0(M)\subset\mathcal{D}(M)$ \\
\hline
Elements & Functions on the base & Differential operators of order zero \\
 & $f(x)$ & $\hat{f}$ \\
\hline
\end{tabular}
\end{center}
\caption{classical versus quantum algebras of functions on the cotangent bundle}
\label{classvsquantalg}
\end{table}

\paragraph{Pseudo-differential operators.} Note that the term ``pseudo-differential operator'' was avoided on purpose, in order to avoid confusion since this technical term is already taken (see \textit{e.g.} \cite{SaintRaymond:1991} for classical textbooks on the subject). 
Roughly, pseudo-differential operators corresponds to functions on phase space with power-law 
asymptotic behaviour (and extra technical requirements). The functional space of pseudo-differential operators seems too small to remain invariant under the action of automorphisms generated by higher-order differential operators.\footnote{Since the idea behind pseudo-differential operators is that they behave asymptotically like differential operators (whose ``order'' can be any real number), they should face the same problem that was encountered for differential operators in Subsection \ref{autom}.}

\subsection{Criteria on the strict product}

\paragraph{Compatibility condition.} On the one hand, the $\mathcal{D}^0(M)$-ring $\mathcal{Q}\mathcal{D}(M)$ of quasi-differential operators is endowed with a non-degenerate character $\hat{\zeta}^*:\mathcal{Q}\mathcal{D}(M)\twoheadrightarrow\mathcal{D}^0(M)$ sending quasi-differential operators $\hat{X}$ to zeroth-order differential operators $\hat{X}_0$ç,. This non-degenerate character extends \eqref{zetatbullet} and is defined exactly in the same way. On the other hand, the $C^\infty(M)$-ring $C^\infty(T^*M)$ is endowed with the degenerate character \eqref{0sectT^*M}.
A strict quantisation is said compatible (with the canonical unit maps and characters) if the following square is commutative:
\be\label{extracondd}
\begin{array}
[c]{ccc}%
C_\star^\infty(T^*M)&\stackrel{\hat{\bullet}}{\longrightarrow}&\mathcal{Q}\mathcal{D}(M)\\
&&\\
\zeta^*\big\downarrow&&\big\downarrow \hat{\zeta}^*\\
&&\\
 C^\infty(M)&\stackrel{\hat{\bullet}}{\longrightarrow}&\mathcal{D}^0(M)
\end{array}
\ee
where the horizontal (respectively, vertical) arrows are isomorphisms (respectively, surjective morphisms) of associative algebras.
Obviously, this requires to start from a compatible quantisations of the Schouten algebra of symbols, since the square \eqref{extracondquant} must be commutative.

\paragraph{Candidate character.} A strict quantisation can be defined equivalently via a non-degenerate character on
the $ C^\infty(M)$-ring $C_\star^\infty(T^*M)$.
The unit map \eqref{canoT^*M} and the degenerate character \eqref{0sectT^*M} of the $C^\infty(M)$-ring $C^\infty(T^*M)$
allow to define how a function $X$ on the cotangent bundle $T^*M$ may act on functions $f$ on the base manifold $M$:
\be\label{definquasidiff}
\hat{X}[f]\,:=\,\zeta^*\Big(\,X\,\star\,\,\tau^*(f)\,\Big)\in C^\infty(M)\quad\text{with}\,\, X\in C^\infty(T^*M)\,\,\text{and}\,\,f\in C^\infty(M)\,.
\ee
which reads, in Darboux coordinates, as $\hat{X}[f]=(X\star f)\,|_{p=0}\,$.
This definition automatically ensures that the square \eqref{extracondd} is commutative, since the constant function $1\in C^\infty(M)$ is the unit element for the strict product: $\hat{X}[1]=\zeta^*(X\star 1)=X_0$\,. However, the corresponding map $\hat{\bullet}:X\mapsto\hat{X}$ need not be an anchor because it may fail to be an algebra morphism.
Nevertheless, the converse statement is true: any compatible strict quantisation is such that the relation \eqref{definquasidiff} holds.\footnote{This can be shown as follows: First, the relation $\hat{X}[f]=(\hat{X}\circ\hat{f})[1]$ is true by the very definition of the map \eqref{hatbull}. Second, the quantisation map is assumed to be an algebra morphism, hence $Q:X\star f\mapsto\widehat{X\star f}=\hat{X}\circ\hat{f}$. Third, the quantisation map is assumed compatible, thus $\hat{X}[f]=\widehat{X\star f}[1]=(X\star f)_0$\,. This ends the proof.
In particular, for functions $X$ on the cotangent bundle which are polynomial in the momenta, the relation \eqref{definquasidiff} reproduces the action on functions $f$ of differential operators $\hat{X}$ with symbol $X$.}

\paragraph{Contact ideal of order zero.} Consider the subalgebra $\ker\zeta^*\subset C^\infty(T^*M)$ spanned by all functions $X$ on the cotangent bundle vanishing on the zero section, \textit{i.e.} such that $X_0=\zeta^*(X)=0$. It is an ideal for both the pointwise product and the Poisson bracket. It will be called the \textit{zeroth-order contact ideal of the cotangent bundle zero-section} and denoted $\mathcal{I}^0\big(\,\zeta(M)\,\big)$. 

\paragraph{Criteria on the strict product.}
The map \eqref{0sectT^*M} induces a character on the $C^\infty(M)$-ring $C_\star^\infty(T^*M)$ iff
the map $\hat{\bullet}:X\mapsto\hat{X}$ defined through \eqref{definquasidiff} is an anchor, which happens iff the underlying strict product satisfies the following condition:
\be\label{strongconsistency}
\forall\, X,Y\in C^\infty(T^*M)\,,\,\,\exists\,Z\in\mathcal{I}^0\big(\,\zeta(M)\,\big)\,:\,X\star Y\,=\,X\star\,\zeta^*(Y)\,+\,Z\,.
\ee
In fact, the map $\hat{\bullet}$ is an anchor of the $C^\infty(M)$-ring $C_\star^\infty(T^*M)$ iff it is a morphism of associative algebras, \textit{i.e.} 
\be\label{identityXYf}
(\hat{Y}\circ\hat{X})[f]=\hat{Y}\big[\,\hat{X}[f]\,\big]\,.
\ee
This condition translates into \eqref{strongconsistency} by using the definition \eqref{definquasidiff}.

\paragraph{Normal-type quantisations.} A strict product $\star$ on the space of functions on the cotangent bundle $T^*M$ such that 
\be \label{normaltype}
f\star X\,=\,f\cdot X\,,\qquad\forall f\in\tau^*\big(\, C^\infty(M)\,\big)\,,\quad\forall X\in C^\infty(T^*M)\,,
\ee 
where $\cdot$ is the pointwise product, will be called of \textit{normal type}.
It is natural to focus on strict products of normal type because the condition \eqref{normaltype} ensures the consistency with the following particular case of the identity \eqref{identityXYf}:
\be\label{indetityfX}
(\hat{f}\circ\hat{X})[g]=f\cdot(\hat{X}[g])\,,
\ee
which holds by the very definition of the map \eqref{hatbull}.
A strict quantisation of the cotangent bundle is said of normal type if the strict quantisation map is $C^\infty(M)$-linear (or, equivalently, if the underlying strict product is of normal type).

\subsection{Strict higher-spin diffeomorphisms}

The automorphisms of the associative algebra $C_\star^\infty(T^*M)$ will be called \textit{strict higher-spin diffeomorphisms of the manifold} $M$. By construction, a strict higher-spin diffeomorphism of $M$ induces a standard diffeomorphism of $M$ iff the commutative algebra $\mathcal{S}^0(M)\cong  C^\infty(M)$ of functions (or, equivalently, the associative algebra $\mathcal{S}(M)$ of symbols)
 is an invariant subspace under the action of this automorphism. This follows immediately from the algebraic interpretation of diffeomorphisms of a manifold as automorphisms of the commutative algebra of functions (\textit{cf.} Subsection \ref{autom}) and the results of \cite{Grabowski:2002,Grabowski:2003b} (\textit{cf.} Subsection \ref{nogoths}).

This notion can be further generalised as follows: an isomorphism $F:C_\star^\infty(T^*M)\stackrel{\sim}{\to}C_\star^\infty(T^*N)$ of associative algebras will be called a \textit{strict higher-spin diffeomorphism from the manifold} $M$ \textit{to the manifold} $N$. One may conjecture that the following three statements are equivalent:
\begin{enumerate}
	\item a strict higher-spin diffeomorphism $C_\star^\infty(T^*M)\stackrel{\sim}{\to}C_\star^\infty(T^*N)$ induces a standard diffeomorphism $M\stackrel{\sim}{\to}N$,
	\item its restriction to the subalgebra of differential operators of order zero is an isomorphism $\mathcal{S}^0(M)\stackrel{\sim}{\to}\mathcal{S}^0(N)$ of commutative algebras, 
	\item its restriction to the whole subalgebra of differential operators is an isomorphism $\mathcal{D}(M)\stackrel{\sim}{\to}\mathcal{D}(N)$ of associative algebras.
\end{enumerate}
The equivalence between the first and second statement is clear, but their equivalence with the third statement remains an open question.

\subsection{Formal completion}

Let us stress that proving the existence of a strict quantisation of the cotangent bundle and/or defining rigorously a suitable completion of the space $\mathcal{D}(M)$ of differential operators are technically challenging problems, that would involve hard-core tools from geometric quantisation and/or functional analysis. In any case, the whole space $C^\infty(T^*M)$ may not be the best suited candidate for the completion one is looking for. The choice of smooth function on the cotangent bundle should just be taken as an indicative example, but this functional space is presumably too big if one looks for ``almost'' differential operators. For instance, it could be replaced with the subspace of $C^\infty(T^*M)$ spanned by functions which are analytic in the momenta. 

Fortunately, for the present purpose there is a simpler way out to avoid any hardcore functional analysis: one may reduce the difficult completion problem to a simpler deformation problem by enlarging even further the functional spaces mentioned above by introducing a formal deformation parameter.

\section{Almost differential operators}\label{almostdiff}

The first subsections contain (some new, but elementary) definitions and lemmas which will be used to introduce in the last subsection an example of formal completion of the almost-commutative algebra of differential operators bypassing the no-go theorem of Grabowski and Poncin.

From now on, the letter $\hbar$ will always denote a formal deformation parameter.\footnote{In the applications to higher-spin gravity, this formal parameter should tentatively be identified with the parameter $\ell$ with the dimension of a length, mentioned below equation \eqref{HX}. But, for the general mathematical considerations of Sections \ref{almostdiff}, \ref{defoquant}, \ref{quasidiffops} and \ref{quotientalgalmostdiffops}, it is a purely formal deformation parameter.}

\subsection{Formal power series over an algebra}\label{formalextalmcom}

\paragraph{Formal power series in $\hbar\,$.} Let $V$ be a complex vector space. Then $V\llbracket\hbar\rrbracket$ denotes the vector space of formal power series in $\hbar$ with coefficients that are elements of $V$. The vector space $V\llbracket\hbar\rrbracket$ is a ${\mathbb{C}}\llbracket\hbar\rrbracket$-module. 

\paragraph{$\hbar$-linearity.} A ${\mathbb{C}}$-linear map $U\in\text{End}\Big(\,V\llbracket\hbar\rrbracket\,\Big)$ is said \textit{$\hbar\,$-linear} if 
\be
U\Big(\,\sum\limits_{n=0}^\infty v_n\hbar^n\,\Big)\,=\,\sum\limits_{n=0}^\infty U(v_n)\,\hbar^n\,,\qquad\forall v_n\in V\,,
\ee
\textit{i.e.} if it is ${\mathbb{C}}\llbracket\hbar\rrbracket$-linear. 
The $\hbar\,$-linear maps span an associative subalgebra of the associative algebra $\text{End}\big(\,V\llbracket\hbar\rrbracket\,\big)$ of endomorphisms of the vector space $V\llbracket\hbar\rrbracket$. Any $\hbar$\,-linear endomorphism $U:V\llbracket\hbar\rrbracket\to V\llbracket\hbar\rrbracket$ is determined uniquely from its restriction $T=U|_V\,:\,V\to V\llbracket\hbar\rrbracket$
to the subspace $V\subset V\llbracket\hbar\rrbracket$ of power series independent of $\hbar$. This restriction is a ${\mathbb{C}}$-linear map from $V$ to $V\llbracket\hbar\rrbracket$, hence it can be thought of as an element of the space $\text{End}(V)\llbracket\hbar\rrbracket$ of formal power series in $\hbar$ with coefficients that are linear maps from the vector space $V$ to itself,   
\be
T\in\text{End}(V)\llbracket\hbar\rrbracket\quad\Leftrightarrow\quad T=\sum\limits_{n=0}^\infty T_n\,\hbar^n\quad\text{with}\quad T_n\in\text{End}(V)\,.
\ee 
Therefore, the $\hbar\,$-linear maps span an associative algebra isomorphic to $\text{End}(V)\llbracket\hbar\rrbracket$. With a slight abuse of notation, the ${\mathbb{C}}$-linear map $T$ and its unique $\hbar$-linear extension $U$ will be denoted by the same symbol from now on.

\paragraph{$\hbar$\,-linear extension of product.} Consider an associative algebra $\mathcal A$\,.  The $\hbar$\,-linear extension of its product endows the vector space $\mathcal{A}\llbracket\hbar\rrbracket$ with a structure of $\mathbb{C}\llbracket\hbar\rrbracket$-algebra. The same is true for a Lie algebra: the $\hbar$\,-linear extension of the bracket of a Lie algebra $\mathfrak{g}$ endows the vector space $\mathfrak{g}\llbracket\hbar\rrbracket$ with a structure of $\mathbb{C}\llbracket\hbar\rrbracket$-Lie algebra. Idem for Poisson algebras.

\paragraph{$\hbar$-filtration.} Consider an associative filtered algebra $\mathcal A$\,. 
Let us denote by $\mathcal{A}\,\llangle\hbar\rrangle$ (respectively, by $\mathcal{A}\,\langle\hbar\rangle$) the vector space spanned by formal power series (respectively, by polynomials) in $\hbar$ 
with coefficients which are of degree \textit{smaller or equal} to the power of $\hbar$, 
\be
a(\hbar)\in \mathcal{A}\,\llangle\hbar\rrangle\quad\Longleftrightarrow\quad a(\hbar)\,=\,\sum\limits_{n=0}^\infty a_n\,\hbar^n\quad\text{with}\quad a_n\in \mathcal{A}_n\,.
\ee
It will be called the \textit{$\hbar$-filtered extension of the 
filtered algebra}.
The $\hbar$\,-linear extension of the product of the associative algebra $\mathcal{A}$ endows the vector space $\mathcal{A}\,\llangle\hbar\rrangle\subset\mathcal{A}\llbracket\hbar\rrbracket$ with a structure of $\mathbb{C}\llbracket\hbar\rrbracket$-subalgebra (indeed $a(\hbar)\, b(\hbar)\in\mathcal{A}\,\llangle\hbar\rrangle$ for $a(\hbar)$ and $b(\hbar)$ in $\mathcal{A}\,\llangle\hbar\rrangle$ because $a_i\,b_j\in\mathcal{A}_{i+j}$ if $a_i\in\mathcal{A}_i$ and $b_j\in\mathcal{A}_j$).
Formal power series in $\hbar$ can be differentiated with respect to the formal variable $\hbar$.
An associative algebra $\mathcal A$ is filtered iff there exists an $\mathbb{C}\llbracket\hbar\rrbracket$-subalgebra $\mathcal{A}\,\llangle\hbar\rrangle\subset\mathcal{A}\llbracket\hbar\rrbracket$ such that $\mathcal{A}\,\llangle\hbar\rrangle\subset\frac{d}{d\hbar}\,\mathcal{A}\,\llangle\hbar\rrangle$\,.
This subalgebra $\mathcal{A}\,\llangle\hbar\rrangle$ defines the filtration of $\mathcal A$ and \textit{vice versa}.

\paragraph{$\hbar$-grading.} Consider an associative graded algebra $\mathcal B$\,. Let us denote by $\mathcal{B}\,\lVert\hbar\rVert$ the associative algebra spanned by formal power series in $\hbar$ with coefficients which are of grading \textit{equal} to the power of $\hbar$.
It will be called the \textit{$\hbar$-graded extension of the associative graded algebra} $\mathcal B$.
The subalgebra $\mathcal{B}\,\lVert\hbar\rVert\subset\mathcal{B}\llbracket\hbar\rrbracket$ is such that $\hbar^n\,\mathcal{B}\,\lVert\hbar\rVert\cap\mathcal{B}\,\lVert\hbar\rVert=\emptyset$ for all $n\in\mathbb{N}_0$. 

\paragraph{$\hbar$-grading associated to an $\hbar$-filtration.} The elements of the form $a(\hbar)=\hbar\,\,b(\hbar)$, where $b(\hbar)\in\mathcal{A}\llbracket\hbar\rrbracket$\,, form a proper ideal $\hbar\,\mathcal{A}\llbracket\hbar\rrbracket\subset\mathcal{A}\llbracket\hbar\rrbracket\,$.
Similarly, the elements of the same form but where $b(\hbar)\in\mathcal{A}\,\llangle\hbar\rrangle$\,, form a proper ideal of $\hbar\,\mathcal{A}\,\llangle\hbar\rrangle\subset\mathcal{A}\,\llangle\hbar\rrangle$\,.
The quotient algebra 
of the $\hbar$-filtered extension $\mathcal{A}\,\llangle\hbar\rrangle$ of the 
filtered associative algebra $\mathcal A$ by the ideal $\hbar\,\mathcal{A}\,\llangle\hbar\rrangle$
is isomorphic to the
$\hbar$-graded extension of the 
associated graded algebra $\mathcal{B}=\text{gr}\mathcal{A}$,
\be\label{quotalg}
\mathcal{A}\,\llangle\hbar\rrangle\,/\,\,\hbar\,\mathcal{A}\,\llangle\hbar\rrangle\,\cong\,\text{gr}\mathcal{A}\,\lVert\hbar\rVert\,.
\ee
An associative filtered algebra $\mathcal{A}$ is almost-commutative iff its $\hbar$-filtered extension $\mathcal{A}\,\llangle\hbar\rrangle$ is commutative modulo $\hbar$, \be
\big[\,\mathcal{A}\,\llangle\hbar\rrangle\,,\,\mathcal{A}\,\llangle\hbar\rrangle\,\big]\subset \hbar\,\mathcal{A}\,\llangle\hbar\rrangle\,.
\ee
which is equivalent to say that the quotient algebra \eqref{quotalg} is commutative.

\paragraph{Refined adjoint representation.} For any associative algebra $\mathcal{A}$, one can form the Lie algebra $\mathfrak{A}$ by endowing the vector space $\mathcal{A}$ with the commutator as Lie bracket.
Consider an almost-commutative algebra $\mathcal{A}$, one can define the following representation of the Lie algebra $\mathfrak{A}\,\llangle\hbar\rrangle$ on the associative algebra $\mathcal{A}\,\llangle\hbar\rrangle$
\be
ad^\hbar\,:\,\mathfrak{A}\,\llangle\hbar\rrangle\to\mathfrak{der}\big(\,\mathcal{A}\,\llangle\hbar\rrangle\,\big)\,:\,b\mapsto ad^\hbar_b
\ee
where 
\be
ad^\hbar_b\,:=\,\frac1{\hbar}\,ad_b
\ee
is an $\hbar\,$-linear derivation of the associative algebra $\mathcal{A}\,\llangle\hbar\rrangle$\,. 

\vspace{3mm}
\noindent{\small\textbf{Technical Lemma (Exponentiation)\,:} Consider an almost-commutative algebra $\mathcal A$. Let us assume that the adjoint action $ad|_{\mathfrak{A}_1}\,:\,\mathfrak{A}_1\to\mathfrak{der}(\mathcal{A})$ of the Lie subalgebra $\mathfrak{A}_1\subset\mathfrak{A}$ of degree one is integrable in the sense that there are one-parameter groups of automorphisms of the almost-commutative algebra $\mathcal A$ with elements of the form $\exp(\,t\, ad_a)\in Aut(\mathcal{A})$ for some $a\in\mathcal{A}_1$. Then the action $ad^\hbar$ of the Lie algebra $\mathfrak{A}\,\llangle\hbar\rrangle$ on the associative algebra $\mathcal{A}\,\llangle\hbar\rrangle$ is essentially integrable, in the sense that all elements $a(\hbar)=\sum_{n=1}^\infty a_n\,\hbar^n\in \mathcal{A}\,\llangle\hbar\rrangle\cap\hbar\,\mathcal{A}\llbracket\hbar\rrbracket$ generate one-parameter groups of automorphisms of the associative algebra $\mathcal A$ of the form $\exp(\,t\, ad^\hbar_a)\in Aut(\mathcal{A}\,\llangle\hbar\rrangle)$ when the coefficient $a_1$ is integrable in the previous sense.
}
\vspace{3mm}

The proof of this technical lemma can be found in appendix.

\subsection{Almost-differential operators}\label{formalalmostdiffops}

A direct corollary of the technical lemma in the previous subsection is that the no-go theorem of Grabowski and Poncin can be bypassed by considering instead the $\hbar$-filtered completion $\mathcal{D}(M)\,\llangle\hbar\rrangle$ of the associative algebra $\mathcal{D}(M)$ of differential operators.
It is spanned by formal power series in $\hbar$ with coefficients that are differential operators on $M$ of order smaller or equal to the power of $\hbar$,
\be\label{Xhbar}
\hat{X}_\hbar\,=\,\sum\limits_{r=0}^\infty\hat{X}_r\,\hbar^r\,,\qquad\hat{X}_r\in\mathcal{D}^r(M)\,.
\ee
They will be called \textit{almost-differential operators} on the manifold $M$.  

\vspace{3mm}
\noindent{\textbf{Yes-go proposition 1 (Almost differential operators)\,:} \textit{An almost-differential operator $\hat{X}_\hbar\in\mathcal{D}(M)\,\llangle\hbar\rrangle\cap\hbar\,\mathcal{D}(M)\llbracket\hbar\rrbracket$ 
is locally integrable to a one-parameter group of automorphisms of the algebra $\mathcal{D}(M)\llangle\hbar\rrangle$ of almost-differential operators.}
}
\vspace{3mm}

More precisely, if $\hat{X}_\hbar\in\mathcal{D}(M)\llangle\hbar\rrangle$ is an almost-differential operator such that $\hat{X}_\hbar|_{\hbar=0}=0$ and if the principal symbol of $\frac{d\hat{X}_\hbar}{d\hbar}|_{\hbar=0}\in\mathcal{D}^1(M)$ is a complete vector field, then $\hat{X}_\hbar$ generates a globally-defined action of the additive group $\mathbb R$ on the algebra $\mathcal{D}(M)\,\llangle\hbar\rrangle$\,.
Note that a similar proposition holds for the Poisson algebra $\mathcal{S}(M)\llangle\hbar\rrangle$
spanned by formal power series in $\hbar$ with coefficients which are symbols of degree smaller or equal to the power of $\hbar$. 

\section{Deformation quantisation: Sample of results}\label{defoquant}

Extending a quantisation of the cotangent bundle to a \textit{strict} quantisation, \textit{i.e.} passing from the space of symbols to the whole space of smooth functions on the cotangent bundle, is not an easy task. It is not even guaranteed to work, a priori. Fortunately, a wealth of positive results are available for a slightly weaker version of this idea: a \textit{deformation} quantisation of the cotangent bundle where one only considers ``formal'' deformations of the pointwise product. Hopefully, one might eventually set the deformation parameter $\hbar$ to be equal to a real number, in which case one would be allowed to speak of a ``strict'' deformation.

References to seminal papers on deformation quantisation and precise location of theorems, quoted without proofs in this section, can be found in the lecture notes by Simone Gutt from which this sample of results has been extracted (\textit{cf.} Sections 3 and 5 of \cite{Gutt:2005}).

\subsection{Star products}

Let $\mathcal M$ be a manifold. A bilinear map 
\be\label{starproddef}
\star\,:\, C^\infty(\mathcal{M})\times C^\infty(\mathcal{M})\to  C^\infty(\mathcal{M})\llbracket\hbar\rrbracket
\ee
can be decomposed as a formal power series 
\be\label{starprodn}
\star=\sum\limits_{n=0}^\infty\star_n\,\hbar^n
\ee
of bilinear maps
\be
\star_n\,:\, C^\infty(\mathcal{M})\times C^\infty(\mathcal{M})\to  C^\infty(\mathcal{M})
\ee
and defines, via $\hbar$\,-(bi)linearity, a bilinear map on the whole space $ C^\infty(\mathcal{M})\llbracket\hbar\rrbracket$.
A bilinear map \eqref{starproddef} defining an associative product on the space $ C^\infty(\mathcal{M})\llbracket\hbar\rrbracket$ and which is a deformation\footnote{One speaks of strict deformations if the series are convergent for $0\leqslant\hbar\leqslant 1$, plus some extra technical assumptions, \textit{cf.} the celebrated definition by Rieffel \cite{Rieffel}.} of the pointwise product on the commutative algebra $C^\infty(\mathcal{M})$ (in the sense that $\star_0=\cdot\,$) is called a \textit{star product on the manifold} $\mathcal M$. The vector space $C^\infty(\mathcal{M})\llbracket\hbar\rrbracket$ endowed with the star product $\star$ is an associative algebra which will be denoted $C_\star^\infty(\mathcal{M})\llbracket\hbar\rrbracket$. 

Any star product on a manifold $\mathcal M$ endows the latter with a structure of Poisson manifold by extracting a Poisson bracket from the leading part of the commutator bracket, \textit{i.e.} 
\be
\{X,Y\}\,:=\,\frac1{\hbar}\,[\,X\,,\,Y\,]_\star\quad\text{modulo}\quad\hbar\,.
\ee
for all $X,Y\in C^\infty(\mathcal{M})$.
Conversely, endowing a Poisson algebra with a star product is called a \textit{deformation of the Poisson algebra of functions}, and one speaks accordingly of a \textit{deformation quantisation of the Poisson manifold}.

A \textit{differential star product} is a star product \eqref{starprodn} such that each bilinear map $\star_n$ is a bidifferential operator on $ C^\infty(\mathcal{M})$. In particular, a \textit{natural star product} is a differential star product such that each bilinear map $\star_n$ is a bidifferential operator of order $n$ in each argument, \textit{i.e.} $\star_n\in (\mathcal{D}^n\otimes\mathcal{D}^n)(\mathcal{M})\llbracket\hbar\rrbracket\,$.

\vspace{3mm}
\noindent{\small\textbf{Example (Normal star product)\,:}
In some Darboux coordinates on the cotangent bundle $T^*M$ of a topologically trivial manifold $M$, 
the normal symbol map \eqref{standardsymbmap} allows to obtain the bilinear map
\be
\stackrel{N}{\star}\,\,:=\,\exp\left(\hbar\,\frac{\overleftarrow{\partial}}{\partial p_{\mu}}\,\frac{\overrightarrow{\partial}}{\partial x^{\mu}}\right)\,,
\label{standardstarprod}
\ee
which provides an example of natural star product of normal type on $T^*M$.
The corresponding set of bidifferential operators reads
\be
X(x,p)\stackrel{N}{\star}_n Y(x,p)\,:=\,\frac1{n!}\,\frac{\partial^n X(x,p)}{\partial p_{\mu_1}\cdots\partial p_{\mu_n}}\,\frac{\partial^n Y(x,p)}{\partial x^{\mu_1}\cdots\partial x^{\mu_n}}\,.
\ee
}


\noindent From now on, one will restrict ourselves to \textit{natural} star products so this assumption will be left implicit.



\subsection{Equivalences of star products and automorphisms of deformations}

The invertible $\hbar$\,-linear maps $T\in\text{End}(V)\llbracket\hbar\rrbracket$ of the form 
\be\label{pertdef}
T=id_V+\sum\limits_{n=1}^\infty T_n\hbar^n\quad\text{with}\quad T_n\in\text{End}(V)
\ee
will be called \textit{perturbative redefinitions of the vector space} $V\llbracket\hbar\rrbracket$.
They form a normal subgroup of the group $GL(\,V\llbracket\hbar\rrbracket\,)\cap \text{End}(V)\llbracket\hbar\rrbracket$ of invertible $\hbar$\,-linear endomorphisms of $V\llbracket\hbar\rrbracket$.

Two star products $\star$ and $\star'$ on the same Poisson manifold $\mathcal{M}$ are \textit{equivalent star products} if there exists a perturbative redefinition $T$ of $C^\infty(\mathcal{M})\llbracket\hbar\rrbracket$ such that 
\be\label{pullbackproduct}
T(f\star g)=(Tf)\star'(Tg)\,,\qquad \forall f,g\in C^\infty(\mathcal{M})\,.
\ee
In other words, an \textit{equivalence of star products} is a perturbative redefinition of the vector space $C^\infty(\mathcal{M})\llbracket\hbar\rrbracket$ which defines an algebra isomorphism between $C_{\star}^\infty(\mathcal{M})\llbracket\hbar\rrbracket$ and $C_{\star^\prime}^\infty(\mathcal{M})\llbracket\hbar\rrbracket$. 
Locally, any two equivalent \textit{differential} star products  $\star$ and $\star'$ on a Poisson manifold $\mathcal M$ are equivalent via a \textit{differential} perturbative redefinition, \textit{i.e.} $T\in\mathcal{D}(\mathcal{M})\llbracket\hbar\rrbracket$ in \eqref{pullbackproduct}.\footnote{This is true globally if the second Betti number of $\mathcal M$ vanishes.}
In general, any associative deformation of a commutative algebra is equivalent to a deformation whose unit element is the same as the one of the undeformed algebra. In this sense, there is no loss of generality in assuming (as will be done from now on) that $1\in C^\infty(\mathcal{M})$ is the unit for both the pointwise product \textit{and} for the star product. 

Let $\mathfrak{C}_\star^\infty(\mathcal{M})\llbracket\hbar\rrbracket$ denote the Lie algebra obtained by endowing the associative $ C_{\star}^\infty(\mathcal{M})\llbracket\hbar\rrbracket$ with the so-called \textit{star commutator} 
\be\label{starcomhbar}
[\,X\,,\,Y\,]_\star\,=\,\hbar\,\{X,Y\}+\mathcal{O}(\hbar^2)
\ee
as Lie bracket. The \textit{star adjoint representation} is the morphism of Lie algebras
\be\label{adjstar}
{}^{\star}ad\,:\,\mathfrak{C}_\star^\infty(\mathcal{M})\llbracket\hbar\rrbracket\,\hookrightarrow\,\mathfrak{inn}\Big(\, C_\star^\infty(\mathcal{M})\llbracket\hbar\rrbracket\,\Big)\,:\,X\mapsto {}^{\star}ad_X
\ee
where
\be
{}^{\star}ad_{X}Y\,:=\,[\,X\,,\,Y\,]_\star
\ee 
is an inner derivation of $C_\star^\infty(\mathcal{M})\llbracket\hbar\rrbracket$

A \textit{self-equivalence of the star product} $\star$ is an equivalence of the star product with itself (\textit{i.e.} $\star=\star'$).
For instance, for any $X\in C_{\star}^\infty(\mathcal{M})\llbracket\hbar\rrbracket$ the automorphism $\exp({}^{\star}ad_X)$ is a self-equivalence called an \textit{inner self-equivalence}. Locally, all self-equivalences of a star product are inner.\footnote{This remains true globally if the first Betti number of $\mathcal M$ vanishes.}  
Two automorphisms of the deformation $C_{\star}^\infty(\mathcal{M})\llbracket\hbar\rrbracket$ which are related by a self-equivalence of the star product $\star$ will be called \textit{equivalent automorphisms}. 

Let $\star$ and $\star'$ be two star products on, respectively, two symplectic manifolds $\mathcal{M}$ and $\mathcal{M}'$. 
Any $\hbar$\,-linear isomorphism  $T: C_{\star}^\infty(\mathcal{M})\llbracket\hbar\rrbracket\stackrel{\sim}{\to} C_{\star'}^\infty(\mathcal{M}')\llbracket\hbar\rrbracket$ of associative algebras is entirely determined by the corresponding formal power series in $\hbar$,
\be\label{Thbar}
T=\sum\limits_{n=0}^\infty T_n\,\hbar^n\quad\text{with}\quad T_n: C^\infty(\mathcal{M})\stackrel{\sim}{\to} C^\infty(\mathcal{M}')\,.
\ee
whose coefficients are linear maps between the Poisson algebras of functions $\mathcal{M}$ and $\mathcal{M}'$. In particular, the term $T_0: C^\infty(\mathcal{M})\stackrel{\sim}{\to} C^\infty(\mathcal{M}')$ at order zero in $\hbar$ is an isomorphism of Poisson algebras, thereby 
defining a symplectomorphism between the two corresponding symplectic manifolds $\mathcal{M}$ and $\mathcal{M}'$. 
Any $\hbar$\,-linear isomorphism is the composition of such a symplectomorphism and an equivalence of star products. The latter will be called the \textit{quantum correction of the symplectomorphism}.

A theorem of Fedosov (see Section 5.5 of his book \cite{Fedosov}) ensures that any symplectomorphism of a symplectic manifold $\mathcal M$ (connected to the identity by a path of symplectomorphisms) can be extended to an $\hbar$\,-linear algebra automorphism of the deformation $C_{\star}^\infty(\mathcal{M})\llbracket\hbar\rrbracket$. For instance, an automorphism of the form $\exp(\frac1{\hbar}{}^{\star}ad_X)$ for a non-vanishing function $X\in C^\infty(\mathcal{M})$ on the symplectic manifold $\mathcal M$ is the deformation of a Hamiltonian symplectomorphism. With a light abuse of terminology, it will be called a \textit{non-trivial inner automorphism of the deformation} $C_{\star}^\infty(\mathcal{M})\llbracket\hbar\rrbracket$. 

\subsection{Derivations}
 
An \textit{$\hbar\,$-linear derivation of the deformation} $ C_{\star}^\infty(\mathcal{M})\llbracket\hbar\rrbracket$ is an $\hbar\,$-linear endomorphism $T$ of the vector space $ C^\infty(\mathcal{M})\llbracket\hbar\rrbracket$ which obeys to the Leibnitz rule with respect to the star product, \textit{i.e.}
\be
T\in\text{End}\Big(\, C^\infty(\mathcal{M})\,\Big)\llbracket\hbar\rrbracket\qquad\text{and}\qquad T(X\star Y)=T(X)\star Y+X\star T(Y)\,,\quad\forall X,Y\in C^\infty(\mathcal{M})\llbracket\hbar\rrbracket.
\ee
The $\hbar$\,-linear derivations span the Lie algebra
\be
\mathfrak{der}\big(\, C_{\star}^\infty(\mathcal{M})\,\big)\cap\text{End}\big(\, C^\infty(\mathcal{M})\,\big)\llbracket\hbar\rrbracket\,.
\ee
Any $\hbar$\,-linear derivation is entirely determined by the corresponding formal power series \eqref{Thbar} in $\hbar$.
In particular, its constant term $T_0$ defines a symplectic vector field on $\mathcal{M}$. 
An $\hbar\,$-linear derivation with non-vanishing constant term will be called a \textit{non-trivial derivation of the deformation} $ C_{\star}^\infty(\mathcal{M})\llbracket\hbar\rrbracket$. The non-trivial derivations are the infinitesimal countepart of $\hbar$\,-linear automorphisms of the deformation $ C_{\star}^\infty(\mathcal{M})\llbracket\hbar\rrbracket$ defining a non-trivial symplectomorphism of $\mathcal{M}$.

The elements $T$ of the associative algebra $\text{End}(V)\llbracket\hbar\rrbracket$ with vanishing constant term, \textit{i.e.} $T_0=0$ in \eqref{pertdef}, will be called \textit{infinitesimal perturbative redefinitions of the vector space} $V\llbracket\hbar\rrbracket$. 
They form the associative ideal $\hbar\,\text{End}(V)\llbracket\hbar\rrbracket$ which will be called the \textit{ideal of infinitesimal perturbative redefinitions of the vector space} $V\llbracket\hbar\rrbracket$,
\be
T\in\hbar\,\text{End}(V)\llbracket\hbar\rrbracket\quad\Leftrightarrow\quad T=\sum\limits_{n=1}^\infty T_n\,\hbar^n\quad\text{with}\quad T_n\in\text{End}(V)\,.
\ee 
The terminology arises from the fact that any perturbative redefinition \eqref{pertdef} is the exponential of an infinitesimal perturbative redefinition. In this sense, the group of perturbative redefinitions can be denoted $\exp\big(\,\hbar\,\text{End}(V)\llbracket\hbar\rrbracket\,\big)$,
\ba
&&T=\exp S\quad\text{with}\quad S\in\hbar\,\text{End}(V)\llbracket\hbar\rrbracket\\
 &\Leftrightarrow& T=id_V+R\quad\text{with}\quad R\in\hbar\,\text{End}(V)\llbracket\hbar\rrbracket\,.
\ea 

Elements which are both $\hbar\,$-linear derivations and infinitesimal perturbative redefinitions of the deformation $ C_{\star}^\infty(\mathcal{M})\llbracket\hbar\rrbracket$ will be called \textit{trivial derivations of the deformation} $C_{\star}^\infty(\mathcal{M})\llbracket\hbar\rrbracket$. They are the infinitesimal counterpart of self-equivalence of star products. 
The Lie algebra
\be\label{trivdefs}
\mathfrak{der}\big(\, C_{\star}^\infty(\mathcal{M})\llbracket\hbar\rrbracket\,\big)\,\cap\,\hbar\,\text{End}\big(\, C^\infty(\mathcal{M})\,\big)\llbracket\hbar\rrbracket
\ee
spanned by trivial derivations is a Lie ideal of the Lie algebra of $\hbar\,$-linear derivations.

An inner derivation of the associative algebra $ C_{\star}^\infty(\mathcal{M})\llbracket\hbar\rrbracket$ is, by definition, the image of an element $X$ by the adjoint representation \eqref{adjstar}, \textit{i.e.}
\be
{}^{\star}ad_{X}\in\mathfrak{inn}\Big(\, C_{\star}^\infty(\mathcal{M})\llbracket\hbar\rrbracket\,\Big)\quad\text{with}\quad X\in C_{\star}^\infty(\mathcal{M})\llbracket\hbar\rrbracket\,.
\ee
Due to the property \eqref{starcomhbar} of the star commutator, any such inner derivation of the associative algebra $ C_{\star}^\infty(\mathcal{M})\llbracket\hbar\rrbracket$ is a trivial derivation. They correspond to infinitesimal inner self-equivalences of the star product $\star$\,.
Accordingly, a derivation of the form
\be\label{genLieder}
\mathcal{L}^\hbar_X\,:=\,\frac1{\hbar}\,\,{}^{\star}ad_{X}\quad\text{with}\quad X\in C^\infty(\mathcal{M})
\ee
will be called a \textit{non-trivial inner derivation of the deformation} $C_{\star}^\infty(\mathcal{M})\llbracket\hbar\rrbracket$.\footnote{This is a slight abuse of terminology since, strictly speaking, they are \textit{not} inner derivations. In fact, by definition ${}^{\star}ad_{X}$ is an inner derivation, but $\mathcal{L}^\hbar_X$ is not (since $1/\hbar\notin{\mathbb{C}}\llbracket\hbar\rrbracket$).}
Locally, all non-vanishing $\hbar$\,-linear derivations of the deformation $C_{\star}^\infty(\mathcal{M})\llbracket\hbar\rrbracket$ are of the form \eqref{genLieder}, but with $X\in C^\infty(\mathcal{M})\llbracket\hbar\rrbracket$ in general.
The non-trivial inner derivation \eqref{genLieder} acts on functions on $\mathcal M$ as follows:
\be
\mathcal{L}^\hbar_X Y\,=\,\{X,Y\}+\mathcal{O}(\hbar)\,,\quad\text{where}\quad X,Y\in C^\infty(\mathcal{M})\,.
\ee
As one can see, the non-trivial inner derivation $\mathcal{L}^\hbar_X$ is a deformation of the Hamiltonian vector field on $\mathcal M$ generated by the Hamiltonian $X$. The non-trivial inner derivations are the infinitesimal contepart of non-trivial inner automorphisms of the deformation $C_{\star}^\infty(\mathcal{M})\llbracket\hbar\rrbracket$.

The table \ref{Sketch} summarises the main $\hbar$-linear transformations of the deformation $C_{\star}^\infty(\mathcal{M})\llbracket\hbar\rrbracket$.

\begin{table}
\begin{center}
\begin{tabular}{
|c|c|c|}
\hline
&&\\
Mathematical objects & \underline{Classical} (undeformed) & Q\underline{uantum} (deformed) \\
&&\\
\hline\hline
&&\\
Un/deformed & Order $\hbar^0$ & Formal power series in $\hbar$\\
&&\\
\hline\hline
Algebra & Symplectic algebra & Associative algebra \\
of functions & $ C^\infty(\mathcal{M})$ &  $ C_\star^\infty(\mathcal{M})\llbracket\hbar\rrbracket$ \\
\hline\hline
Linear maps & $\mathbb{C}$-linear & $\hbar$-linear\\
\hline
Associative algebra of & Endomorphism algebra & Algebra of $\hbar$-linear endomorphisms\\
endomorphisms & $\text{End}\big(\, C^\infty(\mathcal{M}\,\big)$ & $\text{End}\big(\, C^\infty(\mathcal{M})\,\big)\llbracket\hbar\rrbracket$ \\
\hline
Group of & Group of symplectomorphisms & Group of $\hbar$-linear automorphisms\\
finite automorphisms & $\text{Aut}\big(\, C^\infty(\mathcal{M})\,\big)$ & $\text{Aut}\big(\, C^\infty(\mathcal{M})\llbracket\hbar\rrbracket\,\big)\cap\text{End}\big(\, C^\infty(\mathcal{M})\,\big)\llbracket\hbar\rrbracket$ \\
\hline
Lie algebra of & Lie algebra of & Lie algebra of $\hbar$-linear derivations\\
infinitesimal automorphisms & symplectic vector fields & $\mathfrak{der}\big(\, C^\infty(\mathcal{M})\llbracket\hbar\rrbracket\,\big)\cap\text{End}\big(\, C^\infty(\mathcal{M})\,\big)\llbracket\hbar\rrbracket$ \\
\hline
Group of finite & Hamiltonian  & Flows of non-trivial \\
inner automorphisms & flows $\in C^\infty(\mathcal{M})$ & inner automorphisms \\
\hline
Lie algebra of infinitesimal & Lie algebra of & Lie algebra of non-trivial \\
inner automorphisms & Hamiltonian vector fields & inner derivations \\
\hline
Group of & Trivial group & Normal subgroup of pert. redefs\\
finite redefinitions & (identity map) & $\exp\big(\,\hbar\,\text{End}\big(\, C^\infty(\mathcal{M})\,\big)\llbracket\hbar\rrbracket\,\big)$ \\
\hline
Lie algebra of & Trivial algebra & Ideal of infinitesimal pert. redefs\\
infinitesimal redefinitions & (zero map) & $\hbar\,\text{End}\big(\, C^\infty(\mathcal{M}\,\big)\llbracket\hbar\rrbracket$ \\
\hline
Group of & Trivial group & Group of star-product \\
self-equivalences & (identity map) &  self-equivalences \\
\hline
Lie algebra of & Trivial algebra & Lie algebra of \\
infinitesimal self-equivalences & (zero map) & trivial derivations \\
\hline
\end{tabular}
\end{center}
\caption{Deformed vs undeformed algebra of functions}
\label{Sketch}
\end{table}

One should stress that derivations of the deformation $ C_{\star}^\infty(\mathcal{M})\llbracket\hbar\rrbracket$ are power series in $\hbar$ whose general coefficients are \textit{not} derivations of $ C^\infty(\mathcal{M})$:
\be
\mathfrak{der}\Big(\, C_{\star}^\infty(\mathcal{M})\llbracket\hbar\rrbracket\,\Big)\neq\mathfrak{X}(\mathcal{M})\llbracket\hbar\rrbracket\,.
\ee
For instance, non-trivial inner derivations of a deformation $C_{\star}^\infty(\mathcal{M})\llbracket\hbar\rrbracket$ for a differential star product are power series in $\hbar$ whose general coefficients are differential operators on $ C^\infty(\mathcal{M})$:
\be
\mathfrak{inn}\Big(\, C_{\star}^\infty(\mathcal{M})\llbracket\hbar\rrbracket\,\Big)\subset\mathcal{D}(\mathcal{M})\llbracket\hbar\rrbracket\,.
\ee

\vspace{3mm}
\noindent{\small\textbf{Example (Normal star product)\,:} In Darboux coordinates $(x^\mu,p_\nu)$ on the symplectic manifold $\mathcal{M}$, the normal star product \eqref{standardstarprod} allows to calculate from the definition \eqref{genLieder} the explicit form of the non-trivial inner derivations in that case: 
\ba
\mathcal{L}^\hbar_X&=&\sum\limits_{n=1}^\infty\frac{\hbar^{n-1}}{n!}\,\left[\,\frac{\partial^n X(x,p)}{\partial p_{\mu_1}\cdots\partial p_{\mu_n}}\,\frac{\partial^n }{\partial x^{\mu_1}\cdots\partial x^{\mu_n}}\,-\,\frac{\partial^n X(x,p)}{\partial x^{\mu_1}\cdots\partial x^{\mu_n}}\,\frac{\partial^n }{\partial p_{\mu_1}\cdots\partial p_{\mu_n}}\,\right]\,,\nonumber\\
&=&\frac{\partial X}{\partial p_\mu}\,\frac{\partial}{\partial x^\mu}\,-\,\frac{\partial X}{\partial x^\mu}\,\frac{\partial}{\partial p_\mu}\,+\,\mathcal{O}(\hbar)\,.
\ea
}

\section{Quasi differential operators}\label{quasidiffops}

\subsection{Star products of symbols of differential operators}\label{starprodsymbdiffop}

Let $\mathcal{S}(M)[\hbar]$ denote the commutative algebra of polynomials in $\hbar$ with coefficients that are symbols on $M$,
\be\label{comalgpolhbar}
X(x,p\,;\hbar)\,=\,\sum\limits_{r=0}^k X_r(x,p)\,\hbar^r\,,\qquad X_r\in\mathcal{S}(M)\,.
\ee
This commutative algebra is bi-graded: it is graded by the polynomial degree in the momenta and by the polynomial degree in the formal parameter $\hbar$.

Given a quantisation of the cotangent bundle $T^*M$, the vector space $\mathcal{S}(M)$ of symbols is endowed with a structure of almost-commutative algebra via the associative product $\star$ in $\mathcal{S}(M)$ induced from the composition product $\circ$ in $\mathcal{D}(M)$.
Since the Schouten algebra $\mathcal{S}(M)\cong\odot\mathcal{T}(M)$ of symbols is graded by the polynomial degree in the momenta, one may decompose the associative product $\star$ with respect to this grading as in \eqref{hbar1} where 
where $\star_n$ decreases the grading by $n$ (and $\star_0=\cdot$ is the pointwise product).
This allows to define a bilinear map
\be\label{starsymbolhbar}
\star\,:\,\mathcal{S}(M)\times\mathcal{S}(M)\to\mathcal{S}(M)[\hbar]
\ee
as follows
\be\label{starquantisation}
\star=\sum\limits_{n=0}^\infty\,\star_n\,\hbar^n\,.
\ee
The map \eqref{starsymbolhbar} will be loosely called a star product on $\mathcal{S}(M)[\hbar]$. With a slight abuse of notation, the same symbol was used for the induced product and the star product, for the sake of later convenience.

Consider the commutative subalgebra
\be\label{subalgNgrad}
{\mathcal{S}}(M)\,|\hbar|\,:=\,\bigoplus\limits_{n=0}^\infty\,\,\mathcal{S}^n(M)\,\,\hbar^n\,\,\,\subset\,\,\,\mathcal{S}(M)[\hbar]\,,
\ee 
spanned by principal symbols where each momenta comes together with an $\hbar$ factor, \textit{i.e.} its elements take the form
\be
X(x,p\,;\hbar)\,=\,\sum\limits_{r=0}^k \frac1{r!}\,X^{\mu_1\cdots \mu_r}(x)\,p_{\mu_1}\cdots p_{\mu_r}\,\hbar^r\,.
\ee
Equivalently, the coefficient of $\hbar^r$ is a principal symbol of degree $r$\,: $X_r(x,p)=X^{\mu_1\cdots \mu_r}(x)\,p_{\mu_1}\cdots p_{\mu_r}$ in \eqref{comalgpolhbar}.
The commutative algebra ${\mathcal{S}}(M)|\hbar|$ is graded by the polynomial degree in the momenta or, equivalently, in the formal parameter $\hbar$ (since these polynomial degrees coincide). 
The dilatation $p\mapsto \hbar\, p$ of the cotangent spaces by a common scaling factor $\hbar$ defines an isomorphism
\be\label{ihbar}
i:\,\mathcal{S}(M)\,\stackrel{\sim}{\rightarrow}\,\,{\mathcal{S}}(M)|\hbar|\,:\,X(x,p)\mapsto X(x,\hbar\, p)\,.
\ee
of graded algebras, from the algebra $\mathcal{S}(M)$ of symbols, graded by the polynomial degree in momenta, to the $\hbar$-graded algebra $\mathcal{S}(M)|\hbar|$\,.

Consider the associative algebra
\be
\mathcal{D}(M)\,\langle\hbar\rangle\,=\,\bigoplus\limits_{n=0}^\infty\,\mathcal{D}^n(M)\,\,\hbar^n\,\subset\,\mathcal{D}(M)[\hbar]\,,
\ee 
spanned by polynomials in $\hbar$ with components which are differential operators of order smaller or equal to the power of $\hbar$, \textit{i.e.} its elements take the form
\be
\hat{X}_\hbar\,=\,\sum\limits_{r=0}^k \hat{X}_r\,\hbar^r\,,\qquad \hat{X}_r\in\mathcal{D}^r(M)\,.
\ee

The $\hbar$\,-linear extension of the quantisation $q:\mathcal{S}(M)\hookrightarrow\mathcal{D}(M)$ defines a linear injection
\be\label{qbar}
q\,:\,\mathcal{S}(M)|\hbar|\hookrightarrow\mathcal{D}(M)\,\langle\hbar\rangle
\ee
because the restrictions $q_n:\mathcal{S}^n(M)\hookrightarrow\mathcal{D}^n(M)$ of the quantisation are linear injections.
The composition of the canonical isomorphism \eqref{ihbar} with the linear injection \eqref{qbar},
\be\label{Qcirci}
\mathcal{Q}:=Q\circ i\,:\,\mathcal{S}(M)\stackrel{\sim}{\to}\mathcal{D}(M)\,\langle\hbar\rangle
\ee
is a linear injection of the graded vector space $\mathcal{S}(M)$ inside the $\hbar$-filtered vector space $\mathcal{D}(M)\langle\hbar\rangle\,$.
The embedding \eqref{Qcirci} allows to motivate the star product \eqref{starsymbolhbar} on $\mathcal{S}(M)[\hbar]$ as being induced from the composition product $\circ$ in $\mathcal{D}(M)\,\langle\hbar\rangle$, in the sense that
\be
X\star Y\,=\,\mathcal{Q}^{-1}\Big(\,(\mathcal{Q}X)\circ(\mathcal{Q}Y)\,\Big)\,,\qquad\forall X,Y\in\mathcal{S}(M)\,,
\ee
where $\mathcal{Q}$ above is implictly understood as the $\hbar\,$-linear extension of $\mathcal{Q}$.
One can check (by multiplying principal symbols) that the star product is indeed given by the power series \eqref{starquantisation}.

\vspace{3mm}
\noindent{\small\textbf{Example (Normal quantisation)\,:}
One can check explicitly that the bilinear map \eqref{standardstarprod} arises from the normal quantisation \eqref{standardquantisation} by making use of the map
\be\label{qmanormalformal}
Q_N\circ i\,:\,\mathcal{S}(M)\hookrightarrow\mathcal{D}(M)\,\langle\hbar\rangle
\,:\,\sum\limits_{r=0}^k \frac1{r!}\,X^{\mu_1\cdots \mu_r}(x)\,p_{\mu_1}\cdots p_{\mu_r}\,\mapsto\,\sum\limits_{r=0}^k \frac{\hbar^r}{r!}\,X^{\mu_1\cdots \mu_r}(x)\,\partial_{\mu_1}\cdots \partial_{\mu_r}
\ee
which implements the usual quantisation rule $p_\mu\mapsto\hbar\,\partial_\mu$ via the normal ordering prescription.
}

\subsection{Formal quasi-differential operators}

A quantisation of the cotangent bundle $T^*M$ determines a star product on $\mathcal{S}(M)[\hbar]$.
There is a unique deformation quantisation of the cotangent bundle $T^*M$ whose star product on $C^\infty(T^*M)\llbracket\hbar\rrbracket$
\begin{itemize}
	\item[(1)] is differential, and
	\item[(2)] reduces to the 
	star product on $\mathcal{S}(M)[\hbar]$.
\end{itemize}
In fact, since the star product $\star$ is assumed to be differential, it is fixed entirely by its action on symbols, \textit{i.e.} on the subspace $\mathcal{S}(M)\subset C^\infty(T^*M)$. Note that the corresponding star product on the cotangent budle is 
 natural due to our assumption on differential quantisations.  This deformation quantisation of the cotangent bundle (in the sense of Section \ref{defoquant}) will be called the \textit{formal extension} of the quantisation of the cotangent bundle (in the sense of Section \ref{quantisations}).
For instance, the bilinear map \eqref{standardstarprod} provides a formal extension of the normal quantisation \eqref{standardquantisation}.\footnote{
See \cite{Bordemann:1997ep,Bordemann:1997er} for more details on the geometric construction of the homogeneous Fedosov star products on the cotangent bundle generalising the normal 
star product (see also \cite{Pflaum}).} 
Two different quantisations of the cotangent bundle would nevertheless lead to equivalent star products on $\mathcal{S}(M)[\hbar]$ since the composition product in $\mathcal{D}(M)$ remains the same.

Consider a deformation $C_{\star}^\infty(T^*M)\llbracket\hbar\rrbracket$ of the Poisson algebra $C^\infty(T^*M)$ of functions on the cotangent bundle $T^*M$ via a star product $\star$ arising from a formal extension of a quantisation of the cotangent bundle $T^*M$.
Its elements will be called \textit{formal quasi-differential operators on the manifold} $M$.\footnote{As shown in \cite{Bordemann:1997er,Bordemann:2003}, the analogue of the Gelfand–Naimark–Segal construction applies for such formal deformations $C_{\star}^\infty(T^*M)\llbracket\hbar\rrbracket$. This establishes on firm ground the quantum mechanical interpretation of these elements as operators acting on a Hilbert space.
This is not explored here but could be important in the future for looking for a suitable real form of the algebra of formal quasi-differential operators considered here.}
An equivalence class of such deformations $C_{\star}^\infty(T^*M)\llbracket\hbar\rrbracket$ with respect to the equivalence of star products,
will be called a \textit{formal algebra of quasi-differential operators on} $M$, denoted $\mathcal{Q}\mathcal{D}(M)$. Similarly, two such formal algebras of quasi-differential operators over the same manifold $M$ will be considered isomorphic iff their start products are equivalent. By construction, the algebra $\mathcal{D}(M)$ of differential operators is a subalgebra of the formal algebra $\mathcal{Q}\mathcal{D}(M)$ of quasi-differential operators.
A corollary of well-known results in symplectic geometry and deformation quantisation is the uniqueness property of the formal algebra of quasi-differential operators. Strictly speaking, within our technical abilities we were only able to prove it under a mild assumption on the topology of the manifold $M$ but we conjecture that it must be valid for any manifold $M$ and provide some arguments below.

\vspace{3mm}
\noindent{\textbf{Proposition (Uniqueness)\,:} \textit{When the second Betti number of the manifold $M$ vanishes (\,$b_2(M)=0$\,), the formal algebra $\mathcal{Q}\mathcal{D}(M)$ of quasi-differential operators on a manifold $M$ is unique, up to automorphisms.
}}
\vspace{3mm}

\noindent In such case, the algebra $\mathcal{Q}\mathcal{D}(M)$ is entirely determined by $M$, as the terminology and notation suggests ($\hbar$ is absent).

\vspace{3mm}
\noindent{\small\textbf{Proof:} An old theorem of Lichnerowicz \cite{lichnerowicz1979existence} asserts that the differential star product is unique (up to equivalences) on any symplectic manifold $\mathcal M$ whose second Betti number vanishes $b_2(\mathcal{M})=0$. Note that any vector bundle is homotopically equivalent to its base manifold $M$. Therefore, the Betti numbers of the cotangent bundle $T^*M$ are equal to the ones of its base manifold $M$\,: $b_i(T^*M)=b_i(M)$.} $\square$
\vspace{3mm}

\noindent{\small\textbf{Remark:} In the general case (when the second Betti number of the manifold $M$ does not vanish), the vector spaces of equivalence classes of symplectic two-forms and of star products are characterised by the de Rham cohomology in degree two. Nevertheless, one should remember that the star product on $C^\infty(T^*M)\llbracket\hbar\rrbracket$ has been required to reduce to the associative product on $\mathcal{S}(M)[\hbar]$ obtained from a quantisation map $Q:\mathcal{S}(M)\stackrel{\sim}{\to}\mathcal{D}(M)$ of the cotangent bundle. This requirement implies that the symplectic two-form on $T^*M$ is the canonical one. Therefore, one may expect that the admissible star products for the construction of $\mathcal{Q}\mathcal{D}(M)$ are in the same equivalence class as the homogeneous Fedosov star product considered in \cite{Bordemann:1997ep}.\footnote{For the sake of completeness, note that the corresponding star products for a non-canonical symplectic two-form on the cotangent bundle have been studied in \cite{Bordemann:2003}.} Another strategy for proving the uniqueness of the algebra $\mathcal{Q}\mathcal{D}(M)$ would be to show that the star products on $C^\infty(T^*M)\llbracket\hbar\rrbracket$ arising from the formal extension of distinct quantisations of the cotangent bundle $T^*M$ are isomorphic. This is to be expected since two different quantisations of the cotangent bundle lead to equivalent star products on $\mathcal{S}(M)[\hbar]$.}
\vspace{3mm}

The star product of a function $X(x,p)\in  C^\infty(T^*M)$ on the cotangent bundle with a function $f(x)\in C^\infty(M)$ on the base manifold, evaluated at vanishing momenta, produces a formal function on the base manifold. In this sense, formal quasi-differential operators can indeed be interpreted as ``operators'', \textit{i.e.} as linear maps on $ C^\infty(M)\llbracket\hbar\rrbracket$. They are ``formal'' in the sense that they are formal power series in $\hbar$ but, more importantly, they are ``quasi-differential'' in the sense that they can be interpreted as formal power series in $\hbar$ with coefficients which are differential operators.

\vspace{3mm}
\noindent{\textbf{Proposition (Formal quantisation map)\,:} \textit{
Let $Q:\mathcal{S}(M)\stackrel{\sim}{\to}\mathcal{D}(M)$ be a compatible quantisation of the cotangent bundle $T^*M$. There is an $\hbar$-linear surjective anchor $\hat{\bullet}_\hbar:C_\star^\infty(T^*M)\llbracket\hbar\rrbracket\twoheadrightarrow\mathcal{D}(M)\llangle\hbar\rrangle$ of the deformation of the algebra of functions on the cotangent bundle, that extends the compatible quantisation and whose image is the algebra of almost-differential operators.}
\vspace{3mm}

\noindent A map with the above properties will be called a \textit{formal quantisation map},
\be\label{Qmaphbar}
\hat{\bullet}_\hbar\,:\, C_\star^\infty(T^*M)\llbracket\hbar\rrbracket\twoheadrightarrow\mathcal{D}(M)\llangle\hbar\rrangle\,:\,X\mapsto\hat{X}_\hbar\,,
\ee
Concretely, it is defined as
\be\label{Xhbardef}
\hat{X}_\hbar[f]\,:=\,\zeta^*\Big(\,X\,\star\,\,\tau^*(f)\,\Big)\in C^\infty(M)\llbracket\hbar\rrbracket\,.
\ee

\vspace{3mm}
\noindent{\small\textbf{Proof:} Firstly, the map \eqref{Qmaphbar} takes values in $\mathcal{D}\llbracket\hbar\rrbracket$.
Indeed, remember that star products have been assumed to be differential, hence the definition \eqref{Xhbardef} implies that $\hat{X}_\hbar\in\mathcal{D}\llbracket\hbar\rrbracket$. Secondly, the map \eqref{Qmaphbar} is $\hbar$-linear. This is by construction because $\zeta^*$ should be understood in \eqref{Xhbardef} as the $\hbar$-linear extension of \eqref{0sectT^*M}.
Thirdly, the map \eqref{Qmaphbar} is surjective. In fact, the situation is similar to Subsection \ref{starprodsymbdiffop}. The image of a principal symbol $X_r(x,p)\in\mathcal{S}^r(M)$ of degree $r$ in the momenta is a $r$th-order differential operator $\hat{X}_\hbar\in\mathcal{D}^r(M)\,\hbar^r$ of degree $r$ in $\hbar$ whose principal symbol is $X(x,p)\,\hbar^r\,$. Therefore, the images of formal quasi-differential operators of the form $X_r(x,p)\in\mathcal{S}^r(M)\hbar^s$ (for all $r$ and $s$) will span the whole $\mathcal{D}(M)\llangle\hbar\rrangle$. Finally, the map \eqref{Qmaphbar} is a morphism of algebras. Indeed, the definition \eqref{Xhbardef} mimicks \eqref{definquasidiff}. The quantisation of the cotangent bundle is assumed compatible, which means that the quantisation map \eqref{quantisationDMap} is compatible with the definition \eqref{definquasidiff}. Therefore, the $\hbar$-linearity ensures that this remains true for the extension \eqref{Qmaphbar} of \eqref{quantisationDMap}.
}

\vspace{3mm}
\noindent{\small\textbf{Example (Normal quantisation)\,:}
Since
\be
\hat{X}_\hbar[f]\,:=\,\big(\, X(x,p)\,\stackrel{N}{\star} f(x)\,\big)\,\big|_{p=0} = \sum\limits_{n=0}^\infty\frac{\hbar^n}{n!}\,\frac{\partial^n X(x,0)}{\partial p_{\mu_1}\cdots\partial p_{\mu_n}}\,\frac{\partial^n f(x)}{\partial x^{\mu_1}\cdots\partial x^{\mu_n}}\,,
\ee
the normal star product \eqref{standardstarprod} leads to the formal quantisation map
\be\label{Qmaphbarnorm}
\hat{\bullet}^N_\hbar\,:\,X(x,p)\mapsto\hat{X}_\hbar=\sum\limits_{n=0}^\infty\frac{\hbar^n}{n!}\,\frac{\partial^n X(x,0)}{\partial p_{\mu_1}\cdots\partial p_{\mu_n}}\,\partial_{\mu_1}\cdots\partial_{\mu_n}\,\,,
\ee
in agreement with the quantisation map \eqref{qmanormalformal}.
}
\vspace{3mm}

The idea behind the introduction of formal (quasi-)differential operators is that, although the subspace $\mathcal{S}(M)[\hbar]$ (or $C^\infty(T^*M)[\hbar]$\,) of elements polynomial in $\hbar$ reduces to the space of symbols (or of functions on the cotangent bundle) if one would set the formal variable to $\hbar=1$, this is not true for the space $C^\infty(T^*M)\llbracket\hbar\rrbracket$ of formal quasi-differential operators because the evaluation at $\hbar=1$ can be divergent. 
Nevertheless, one can interpret the formal algebra $\mathcal{Q}\mathcal{D}(M)$ of quasi-differential operators as a completion of the almost-commutative algebra $\mathcal{D}(M)$ of differential operators. Metaphorically it would produce the desired algebra 
of strict quasi-differential operators, if one were allowed to set $\hbar=1$ in the formulae. This will be good enough for our purpose. 

\section{Higher-spin diffeomorphisms}\label{HSdiffsfinally}

Let us stress that if $\hbar$ is treated as a real parameter (rather than a formal variable), then it becomes natural to consider only isomorphisms and automorphisms that are $\hbar$\,-linear.

\subsection{Looking for higher-spin diffeomorphisms}

The known general results on star product (and their equivalence) ensure that deformation quantisation provides a possible cure of the formal exponentiation of higher-spin Lie derivatives. The Appendix B of \cite{Bordemann:1997er} provides a very concise review of the mathematically rigorous results on Heisenberg-picture time evolution of operators in deformation quantisation.

One can define the higher-spin Lie derivative of a differential operator $\hat{Y}(x,p)\in\mathcal{D}(M)$ along a differential operator $\hat{X}\in\mathcal{D}(M)$ as a trivial inner derivation of the deformation $C_{\star}^\infty(T^*M)\llbracket\hbar\rrbracket$ associated to the corresponding symbols $X,Y\in\mathcal{S}(M)$. This defines the one-parameter group of inner self-equivalences:
\ba
Y(x,p)&\mapsto& Y_t(x,p\,;\hbar)
=\exp\Big(\,t\,\,{}^{\star}ad_{X}\,\Big)\,Y(x,p)\label{1pargroupautoo}\\
&&=\exp_\star\big(\,+t\,X(x,p)\,\big)\,\star\, Y(x,p)\,\,\star\,\,\exp_\star\big(\,-t\,X(x,p)\,\big)\,.\nonumber
\ea
where 
\be
\exp_\star A\,:=\,\sum\limits_{n=0}^\infty\frac1{n!}\underbrace{A\star\cdots\star A}_{n\,\,\text{factors}}\,.
\ee
It remains true that if $X(x,p)\in\mathcal{S}^r(M)$ is a symbol of degree $r>1$ then its star adjoint action on the algebra of symbols
\be
{}^{\star}ad_{X}\,:\,\mathcal{S}^q(M)\to\mathcal{S}^{q+r-1}(M)[\hbar]
\ee
increases the degree in momenta by $r-1>0$. However the difference now is that each such higher-spin Lie derivative brings at least one power of $\hbar$. Therefore, 
\be
Y_t(x,p\,;\hbar)\in\mathcal{S}(M)\llbracket\hbar\rrbracket\qquad\text{if}\qquad Y(x,p)\in\mathcal{S}(M)\,.
\ee
Consequently, inner self-equivalences generated by higher-order differential operators $\hat{X}\in\mathcal{D}(M)$ are inner automorphisms of the subalgebra $\mathcal{S}_\star(M)\llbracket\hbar\rrbracket\subset C_{\star}^\infty(T^*M)\llbracket\hbar\rrbracket$ denoting the space of symbols $\mathcal{S}(M)$ endowed with the star product. Unfortunately, these inner automorphisms are trivial automorphisms of $\mathcal{S}_\star(M)\llbracket\hbar\rrbracket$. In fact, their classical limit is the identity: 
\be
Y_t(x,p\,;\hbar\to 0)=Y(x,p)
\ee
for all $t\in\mathbb R$. This is consistent with the Grabowski-Poncin no-go theorem.

A proper refinement of the previous attempt is to consider instead the mapping:
\be\label{1pargroupautooo}
Y(x,p)\mapsto Y^\hbar_t(x,p)\,:=\exp\Big(\,t\,\mathcal{L}^\hbar_{X}\,\Big)\,Y(x,p)\,.
\ee
where $\mathcal{L}^\hbar_{X}$ denotes a non-trivial inner derivation\eqref{genLieder} of the deformation $C_{\star}^\infty(T^*M)\llbracket\hbar\rrbracket$.
More generally, a derivation of the form \eqref{genLieder} with $X\in C_{\star}^\infty(T^*M)\llbracket\hbar\rrbracket$ will be called a \textit{formal higher-spin Lie derivative along a quasi-differential operator}. Locally, any $\hbar$\,-linear derivation of the deformation $ C_{\star}^\infty(T^*M)\llbracket\hbar\rrbracket$ is a formal higher-spin Lie derivative.
The classical limit of \eqref{1pargroupautooo} is the symplectomorphism
\be\label{1pargroupautooosymplectomorphism}
Y(x,p)\mapsto Y^{\hbar\to 0}_t(x,p)=\exp\Big(\,t\,\{\,X(x,p)\,,\,\,\,\}\,\Big)\,Y(x,p)
\ee
generated by the Hamiltonian vector field with $X(x,p)$ as Hamiltonian.  Let us stress again that the subspace $\mathcal{S}(M)\subset C^\infty(T^*M)$ of symbols is \textit{not} preserved  by such symplectomorphisms, consistently with the Grabowski-Poncin no-go theorem.
However, the transformations \eqref{1pargroupautooosymplectomorphism} are well-defined on the whole Poisson algebra $C^\infty(T^*M)$. In other words, generically $Y^{\hbar\to 0}_t\in C^\infty(T^*M)$ even when $X,Y\in\mathcal{S}(M)$.

\subsection{Formal higher-spin diffeomorphisms}

An $\hbar$\,-linear isomorphism of associative algebras between two deformations $C_\star^\infty(T^*M)\llbracket\hbar\rrbracket$ and $C_{\star'}^\infty(T^*M')\llbracket\hbar\rrbracket$ will be called a \textit{formal higher-spin diffeomorphism between the manifold} $M$ \textit{and the manifold} $M'$.

A corollary of the known results for star products on symplectic manifolds (see Proposition 9.4 of \cite{Gutt}) is that any formal higher-spin diffeomorphism $\Phi:C_\star^\infty(T^*M)\llbracket\hbar\rrbracket\stackrel{\sim}{\to} C_{\star'}^\infty(T^*M')\llbracket\hbar\rrbracket$ between $M$ and $M'$ is the composition $\Phi= F^*\circ T$ of its classical limit $F^*:C_\star^\infty(T^*M)\stackrel{\sim}{\to} C_{\star'}^\infty(T^*M')$ (\textit{i.e.} an isomorphism of Poisson algebras, associated to a symplectomorphism $F:T^*M'\stackrel{\sim}{\to}T^*M$ between the corresponding cotangent bundles) 
and a quantum correction $T:C_\star^\infty(T^*M)\stackrel{\sim}{\to}C_\star^\infty(T^*M)$ (\textit{i.e.} a self-equivalence of star product).

Therefore, there is a one-to-one correspondence between:
\begin{enumerate}
  \item symplectomorphisms from the cotangent bundle $T^*M'$ to the cotangent bundle $T^*M$,	
	\item classical limit of formal higher-spin diffeomorphisms between the manifolds $M$ and $M'$,
	\item equivalence classes of $\hbar$\,-linear isomorphisms between the associative algebras $C_\star^\infty(T^*M)\llbracket\hbar\rrbracket$ and $C_{\star'}^\infty(T^*M')\llbracket\hbar\rrbracket$ of formal quasi-differential operators with respect to the equivalence of star products,
	\item $\hbar$\,-linear isomorphisms between the formal algebras $\mathcal{Q}\mathcal{D}(M)$ and $\mathcal{Q}\mathcal{D}(M')$ of quasi-differential operators.
\end{enumerate}

\subsection{Formal higher-spin flows}

An $\hbar$\,-linear algebra automorphism of the deformation $C_\star^\infty(T^*M)\llbracket\hbar\rrbracket$ will be called a \textit{formal higher-spin diffeomorphism of the manifold} $M$. They form a group which will be denoted $HSDiff(M)$. 
The formal higher-spin diffeomorphisms exhaust all automorphisms of the associative algebra $C_\star^\infty(T^*M)\llbracket\hbar\rrbracket$ of formal quasi-differential operators, up to spurious perturbative redefinitions of $\hbar$. Locally, any non-trivial formal higher-spin diffeomorphism of $M$ is a non-trivial inner automorphism of $C_\star^\infty(T^*M)\llbracket\hbar\rrbracket$.

Any symplectomorphism of the cotangent bundle $T^*M$, connected to the identity by a path of symplectomorphisms, admits an extension to a formal higher-spin diffeomorphisms of $M$. Moreover, this extension is (locally) unique up to self-equivalences. In fact, formal higher-spin diffeomorphisms on a manifold $M$ can be thought of as ``quantum-corrected'' symplectomorphisms on the cotangent bundle $T^*M$.

An action of the additive group $\mathbb R$ on the deformation $C^\infty(T^*M)\llbracket\hbar\rrbracket$ will be called a \textit{(globally-defined) formal higher-spin flow on the manifold} $M$.
One will also admit locally-defined formal higher-spin flows, \textit{i.e.} actions of a Lie subgroup $I\subset \mathbb R$ of the additive group $\mathbb R$ on the associative algebra $C^\infty(T^*N)\llbracket\hbar\rrbracket$ of formal differential operators on a submanifold $N\subset M$.
This allows to formulate a solution to the main problem addressed in this paper.  

\vspace{3mm}
\noindent{\textbf{Yes-go proposition 2 (Formal quasi-differential operator)\,:} 
\textit{Any formal quasi-differential operator $X\in C_\star^\infty(T^*M)\llbracket\hbar\rrbracket$ is integrable to a formal higher-spin flow on $M$,
\textit{i.e.} a group morphism  
\be
\exp(\bullet\mathcal{L}^{\hbar}_X)\,:\,I\to HSDiff(N)\,:\,t\mapsto\exp(\,t\,\mathcal{L}^{\hbar}_X)\,,
\ee
defined for an open subset $I\subseteq\mathbb{R}$ (\textit{e.g.} an open interval $I=]a,b[$\,) and submanifold $N\subseteq M$.
}}
\vspace{3mm}

More precisely, the formal higher-spin flow is the combination
\be
\exp(\,t\,\mathcal{L}^{\hbar}_X)=T_t\circ\exp(\,t\,\mathcal{L}^{\hbar\to 0}_X)
\ee
of the corresponding Hamiltonian flow on $T^*M$, \textit{i.e.} the group morphism 
\be
\exp(\bullet\mathcal{L}^{\hbar\to 0}_X)\,:\,I\to Diff(T^*N)\,:\,t\mapsto\exp(\,t\,\mathcal{L}^{\hbar\to 0}_X)\,,
\ee 
with the quantum correction map
\be
T_\bullet\,:\,{\mathbb R}\to \exp\big(\,\hbar\,\mathcal{D}(M)\llbracket\hbar\rrbracket\,\big)\,:\,t\mapsto T_t
\ee
where the perturbative redefinitions 
\be
T_t=id+\sum\limits_{n=1}^\infty \stackrel{(n)}{T}_t\hbar^n\quad\text{with}\quad \stackrel{(n)}{T}_t\in\mathcal{D}^{2n}(T^*M)\,,
\ee
have coefficients which are differential operators on the cotangent bundle $T^*M$ of order equal to twice the corresponding degree in $\hbar$.\footnote{This last property holds because the star product is natural. Detailed statements about the quantum correction to the Heisenberg-picture time evolution can be found in Appendix B of \cite{Bordemann:1997er}.} 

\vspace{3mm} 
\noindent{\small\textbf{Proof:} Any formal quasi-differential operator $X\in C_\star^\infty(T^*M)\llbracket\hbar\rrbracket$ defines a formal higher-spin Lie derivative \eqref{genLieder} along $X$ whose classical limit $\mathcal{L}^{\hbar\to 0}_{X}$ is a Hamiltonian vector field on $T^*M$ for the Hamiltonian $X|_{\hbar=0}\in C^\infty(T^*M)$. As any vector field on a manifold, this vector field is integrable to a local flow. In particular, the Hamiltonian vector field $\mathcal{L}^{\hbar\to 0}_{X}$ on $T^*M$ is integrable to a local Hamiltonian flow $\exp(\bullet\mathcal{L}^{\hbar\to 0}_{X})$ on $T^*M$, defined for an open subset $I\subseteq\mathbb{R}$ and submanifold $N\subseteq M$.
A theorem of Fedosov (Proposition 5.5.6 of \cite{Fedosov}) ensures that each symplectomorphisms $\exp(\,t\,\mathcal{L}^{\hbar\to 0}_{X})$ for $t\in I$ can be extended to a formal higher-spin diffeomorphism of $M$, which can be denoted $\exp(\,t\,\mathcal{L}^{\hbar}_X)$.
} $\square$ 

\section{Quotient algebra of almost-differential operators}\label{quotientalgalmostdiffops}

As was shown in Section \ref{HSdiffsfinally}, the algebra of (formal) quasi-differential operators actually meets our goal in that it allows to define (formal) higher-spin diffeomorphisms. Nevertheless, this completion of the algebra of differential operators is quite huge. Two smaller algebras are actually available: one is a subalgebra and one is a quotient algebra of the algebra of formal quasi-differential operators.

\subsection{Subalgebra}

There is an algebra sitting in between the two algebras $\mathcal{S}_\star(M)\llbracket\hbar\rrbracket$ and $C_{\star}^\infty(T^*M)\llbracket\hbar\rrbracket$: it is spanned by formal power series in $\hbar$ whose coefficients are smooth functions of the base $M$ of the cotangent bundle $T^*M$ but analytic functions of the momenta (see \cite{Bordemann:1997ep} for a proof). 
This small completion is of interest. However, it is another one that we will investigate here.

\subsection{Quotient algebra}

\paragraph{Infinite-order contact ideal of the zero section.} Consider the Poisson subalgebra $\mathcal{I}^\infty\big(\,\zeta(M)\,\big)\subset C^\infty(T^*M)$ spanned by all functions on the cotangent bundle vanishing on the zero section together with all their derivatives along momenta. More concretely, all derivatives along momenta vanish when evaluated at zero momenta:\footnote{For instance, if the base manifold is the Euclidean space $M={\mathbb R}^n$, then the function $f(x)\exp(-1/\vec{p}\,{}^2)$ is such that its Taylor series along the direction of momenta vanishes on the zero section $\vec{p}=\vec{0}$.} 
\be
X(x,p)\in\mathcal{I}^\infty\big(\,\zeta(M)\,\big)\quad\Longleftrightarrow\quad X(x,p)\in C^\infty(T^*M)\quad\text{and}\quad\frac{\partial X(x,0)}{\partial p_{\mu_1}\cdots\partial p_{\mu_k}}=0\,,\quad\forall k\in{\mathbb N}\,.
\ee
The subalgebra $\mathcal{I}^\infty\big(\,\zeta(M)\,\big)$ is an ideal of $ C^\infty(T^*M)$ for both the pointwise product and the Poisson bracket. It will be called the \textit{infinite-order contact ideal of the zero section  of the cotangent bundle}.
Note that this ideal does not contain any non-trivial symbol, \textit{i.e.}
\be
\mathcal{I}^\infty\big(\,\zeta(M)\,\big)\,\cap\,\mathcal{S}(M)\,=\,\{0\}
\ee

\paragraph{Infinitesimal neighborhood of the zero-section.} The quotient 
\be
\mathcal{J}^\infty\big(\,\zeta(M)\,\big)\,:=\, C^\infty\big(\,T^*M\,\big)\,/\,\mathcal{I}^\infty\big(\,\zeta(M)\,\big)
\ee
of the Poisson algebra of functions the cotangent bundle $T^*M$ by the infinite-order contact ideal of the zero section $\zeta(M)\subset T^*M$, is a Poisson algebra, whose elements will be called \textit{jet fields on the infinitesimal neighborhood of the cotangent bundle zero-section}. They can be thought of as Taylor series at the origin of cotangent spaces (\textit{i.e.} at zero momenta) of smooth functions on the cotangent bundle:
\be\label{formalTaylorserieszerosection}
X(x\,;p)=\sum\limits_{r=0}^\infty X^{\mu_1\cdots\mu_r}(x)\,p_{\mu_1}\cdots p_{\mu_1}\,.
\ee

\paragraph{Deformation quantisation.} For any differential star product $\star\,$, the subspace $\mathcal{I}^\infty\big(\,\zeta(M)\,\big)\llbracket\hbar\rrbracket\subset C^\infty(T^*M)\llbracket\hbar\rrbracket$ is an ideal of the deformed algebra, \textit{i.e.} with respect to the star product. As such, it will be denoted $\mathcal{I}_\star^\infty\big(\,\zeta(M)\,\big)\llbracket\hbar\rrbracket\subset C_{\star}^\infty(T^*M)\llbracket\hbar\rrbracket$. 
The quotient 
\be\label{quotientJ}
 \mathcal{J}^\infty_\star\big(\,\zeta(M)\,\big)\llbracket\hbar\rrbracket
\,:=\, C_\star^\infty(T^*M)\llbracket\hbar\rrbracket\,/\,\mathcal{I}_\star^\infty\big(\,\zeta(M)\,\big)\llbracket\hbar\rrbracket
\ee
of the deformation $C_\star^\infty(T^*M)\llbracket\hbar\rrbracket$ by the ideal $\mathcal{I}_\star^\infty\big(\,\zeta(M)\,\big)\llbracket\hbar\rrbracket$ is an associative algebra, which is a deformation of the Poisson algebra $\mathcal{J}^\infty\big(\,\zeta(M)\,\big)$ of jet fields on the infinitesimal neighborhood of the cotangent bundle zero-section. 
One may interpret this construction as a deformation quantisation of the infinitesimal neighborhood of the cotangent bundle zero-section. 

\paragraph{Injective anchor.} The formal quantisation map $\hat{\bullet}_\hbar$ in \eqref{Qmaphbar} is surjective but not injective.
In fact, it is not injective because it has a non-trivial kernel. The kernel of $\hat{\bullet}_\hbar$ is precisely the $\infty$-contact ideal $\mathcal{I}^\infty\big(\zeta(M)\big)\llbracket\hbar\rrbracket$ of the zero section of the cotangent bundle.\footnote{This can be checked explicitly for the normal star product in Darboux coordinates, via the formula \eqref{Qmaphbarnorm}. Nevertheless, the conclusion is coordinate-free and remains valid for any equivalent differential star product.} Therefore, the quotient map
\be\label{isomformalalmostdiff}
\hat{\bullet}_\hbar\,:\, \mathcal{J}^\infty_\star\big(\,\zeta(M)\,\big)\llbracket\hbar\rrbracket\stackrel{\sim}{\to}\mathcal{D}(M)\,\llangle\hbar\rrangle\,:\,X(x\,;p,\hbar)\mapsto\hat{X}_\hbar(x\,;\partial)\,\,.
\ee
is an isomorphism of associative algebras between $\mathcal{J}^\infty_\star\big(\,\zeta(M)\,\big)\llbracket\hbar\rrbracket$ and the algebra $\mathcal{D}(M)\llangle\hbar\rrangle$. Therefore, the elements of $\mathcal{J}^\infty_\star\big(\,\zeta(M)\,\big)\llbracket\hbar\rrbracket$ can be thought as almost-differential operators (although the realisation is rather different).

\paragraph{Example (Normal quantisation).}
The formal quantisation map \eqref{Qmaphbarnorm} takes exactly the same form if one replaces the functions $X(x,p)$ on the cotangent bundle by Taylor series \eqref{formalTaylorserieszerosection} at the zero section. The analogue of the map \eqref{qmanormalformal} would be the following embedding
\ba\label{qmanormalformal'}
Q_N\circ i&:& \mathcal{J}^\infty_\star\big(\,\zeta(M)\,\big)\hookrightarrow\mathcal{D}(M)\,\llangle\hbar\rrangle\\
&:&\sum\limits_{r=0}^\infty \frac1{r!}\,X^{\mu_1\cdots \mu_r}(x)\,p_{\mu_1}\cdots p_{\mu_r}\,\mapsto\,\sum\limits_{r=0}^\infty \frac{\hbar^r}{r!}\,X^{\mu_1\cdots \mu_r}(x)\,\partial_{\mu_1}\cdots \partial_{\mu_r}\,,
\ea
whose $\hbar\,$-linear extension would reproduce the isomorphism $\hat{\bullet}^N_\hbar: \mathcal{J}^\infty_\star\big(\,\zeta(M)\,\big)\llbracket\hbar\rrbracket\stackrel{\sim}{\to}\mathcal{D}(M)\llangle\hbar\rrangle$\,. Its inverse is the map
\ba\label{standardsymbmap'}
(\hat{\bullet}^N_\hbar)^{-1}&:&\mathcal{D}(M)\llangle\hbar\rrangle\stackrel{\sim}{\to} \mathcal{J}^\infty_\star\big(\,\zeta(M)\,\big)\llbracket\hbar\rrbracket\\
&:&\hat{X}_\hbar\mapsto X(x\,;p,\hbar)\,=\,\exp(-\tfrac{1}{\hbar}\,p_\mu x^\mu)\,\hat{X}[\,\exp(\tfrac{1}{\hbar}\,p_\mu x^\mu)\,]
\ea
 
\paragraph{Quotient algebra of almost differential operators.}
The algebra 
\be\label{isomorphismDC}
\mathcal{D}(M)\,\llangle\hbar\rrangle\cong \mathcal{J}^\infty_\star\big(\,\zeta(M)\,\big)\llbracket\hbar\rrbracket
\ee
of almost-differential operators is a completion with a reasonable size (since the quotient throws away all elements which are invisible in the operatorial interpretation) of the algebra $\mathcal{D}(M)$ of differential operators. 
Unfortunately, the deformation $\mathcal{J}^\infty_\star\big(\,\zeta(M)\,\big)\llbracket\hbar\rrbracket$ of the Poisson algebra $\mathcal{J}^\infty\big(\,\zeta(M)\,\big)$ of jet fields on the infinitesimal neighborhood of the cotangent bundle zero-section is constructed as a quotient (\textit{cf.} \eqref{quotientJ}\,) and, as such, requires more care than the algebra $\mathcal{D}(M)\llangle\hbar\rrangle$ of quasi-differential operators. For instance, the zero section $\zeta(M)$ is \textit{not} preserved by generic symplectomorphisms of the cotangent bundle, so the quotient algebra $\mathcal{J}^\infty_\star\big(\,\zeta(M)\,\big)\llbracket\hbar\rrbracket$ would meet the same conceptual problem as the Schouten algebra $\mathcal{S}(M)$ of symbols, as far as automorphisms are concerned.
This implies that only those formal higher-spin diffeomorphisms on $M$ whose classical limit are symplectomorphism of $T^*M$ sending the zero section $\zeta(M)\subset T^*M$ into itself will descend from automorphism of the algebra $ C_\star^\infty\big(\,T^*M\,\big)\llbracket\hbar\rrbracket$ of quasi-differential operators to automorphisms of the algebra $\mathcal{J}^\infty_\star\big(\,\zeta(M)\,\big)\llbracket\hbar\rrbracket$ of almost-differential operators.\footnote{From the point of view of higher-spin gravity, the status of such a strong condition on higher-spin diffeomorphisms is unclear since it would remove the Maxwell gauge symmetries of the spin-one sector (since they correspond to vertical automorphisms of the cotangent bundle that do \textit{not} preserve the zero section). Nevertheless, they appear as a reasonable candidate subclass of higher-spin symmetries which could be compatible with some weak notion of locality.}

\paragraph{Zeroth-order contact ideal of the zero section.}
The zeroth-order contact ideal $\mathcal{I}^0\big(\,\zeta(M)\,\big)=\text{Ker}\,\zeta^*\subset C^\infty(T^*M)$ of the cotangent bundle zero-section is spanned by all functions $X$ on the cotangent bundle vanishing on the zero section, \textit{i.e.} such that $X_0=\zeta^*(X)=0$. It is an ideal for both the pointwise product and the Poisson bracket. 
The infinite-order contact ideal $\mathcal{I}^\infty\big(\,\zeta(M)\,\big)\subset\mathcal{I}^0\big(\,\zeta(M)\,\big)$ remains an ideal inside the zeroth-order one.
The quotient algebra
\be
\mathcal{I}\big(\,\zeta(M)\,\big)\,:=\,\mathcal{I}^0\big(\,\zeta(M)\,\big)\,/\,\mathcal{I}^\infty\big(\,\zeta(M)\,\big)
\ee
of the zeroth-order ideal of the zero section $\zeta(M)\subset T^*M$ by the infinite-order one, is a Poisson algebra. Its elements can be thought of as Taylor series \eqref{formalTaylorserieszerosection} vanishing at zero momenta (\textit{i.e.} the sum starts at $r=1$).
Due to the isomorphism \eqref{isomorphismDC}, a corollary of the yes-go proposition in Subsection \ref{formalalmostdiffops} is that any element $X\in\mathcal{I}\big(\,\zeta(M)\,\big)\llangle\hbar\rrangle$ is locally integrable to a one-parameter group of non-trivial automorphisms $\exp(-t\,\mathcal{L}^\hbar_{X}\,)$ of the algebra $\mathcal{J}^\infty_\star\big(\,\zeta(M)\,\big)[[\hbar]]$ of almost-differential operators.

\section{Conclusion}\label{conclusion}

General results in deformation quantisation provide cures of the obstruction to the integration of differential operators to one-parameter groups of inner automorphisms. One does so by considering some larger algebras (which can be thought of as a sort of completion of the algebra of differential operators) on which ``formal'' higher-spin diffeomorphisms are well-defined. The latter are nothing but a fancy name for quantum symplectomorphisms addressed in the realm of deformation quantisation.

The obstruction to the integration of higher-degree Hamiltonian vector fields to one-parameter groups of inner automorphisms of the Schouten algebra of symbols was bypassed in Section \ref{symplectomcotgt} by enlarging the latter to the Poisson algebra of functions on the cotangent bundle. Similarly, the obstruction to the integration of higher-spin Lie derivatives to one-parameter groups of inner automorphisms of the almost-commutative algebra of differential operators was bypassed in Section \ref{HSdiffsfinally} by enlarging the latter to the associative algebra of formal quasi-differential operators. Moreover, by properly taking into account the equivalence relations underlying this extension, there is a one-to-one correspondence between the symmetries of their classical limits, \textit{i.e.} between symplectomorphisms of the cotangent bundle and the classical limit of formal higher-spin diffeomorphisms.
 
Overcoming the obstacle was done at the price of considering \textit{formal} deformations. It would be nice to see if similar results hold for some associative algebra of \textit{strict} quasi-differential operators. Nevertheless, the results obtained from deformation quantisation will be taken as an indication that strict higher-spin diffeomorphisms can be defined rigorously. The table \ref{sketch2} summarises and compares the symmetries of the classical vs quantum tangent bundle. As a sign of optimism, no explicit distinction was made between ``strict'' and ``formal'' in the table.

\begin{table}
\begin{center}
\small
\begin{tabular}{
|c|c|c|}
\hline
& Classical & Quantum \\
\hline\hline
Algebra & Poisson algebra (symplectic) & Associative algebra (central) \\
 & $C^\infty(T^*M)$ &  $\mathcal{Q}\mathcal{D}(M)$ \\
\hline
Elements & Functions on the cotangent bundle & Quasi-differential operators \\
 & $X(x,p)$ & $\hat{X}(x,\partial)$ \\
\hline\hline
Finite & Symplectomorphisms & Higher-spin \\
automorphisms & of $T^*M$ &  diffeomorphisms of $M$ \\
\hline
Flow of inner & Hamiltonian flow & Higher-spin flow \\
automorphisms & on $T^*M$ & on $M$\\
\hline\hline
Infinitesimal & Symplectic vector field & Infinitesimal higher-spin \\
automorphism &  on $T^*M$ &  diffeomorphism of $M$\\
\hline
Inner & Hamiltonian vector & Higher-spin \\
derivation & field on $T^*M$ & Lie derivative on $M$\\
\hline
\end{tabular}
\end{center}
\caption{Automorphisms of classical versus quantum algebras of functions on the cotangent bundle}
\label{sketch2}
\end{table}

\section*{Acknowledgments}

Thomas Basile, Kevin Morand and Paolo Saracco are thanked for useful discussions. The author also thank the Erwin Schr\"{o}dinger International Institute for Mathematics and Physics (Vienna) for hospitality during the thematic programme ``Geometry for Higher Spin Gravity: Conformal Structures, PDEs, and Q-manifolds'' (August 23rd — September 17th, 2021) during which the final version of this work was completed.

\appendix

\section*{Proof of technical lemma}\label{prooftechnicalemma}

One possible strategy for the proof of the ``technical lemma'' in Subsection \ref{formalextalmcom} is to start by showing that the lemma holds in the particular case of homogeneous elements $a(\hbar)=a_n\,\hbar^n$ for all $n\geqslant 1$, and then to conclude the proof by showing that the corresponding inner automorphisms generate the generic case via composition. The proof is organised slightly differently but follows the same logic.

Firstly, it is clear that the lemma holds in the particular case $a(\hbar)=a_1\,\hbar$ with $a_1\in\mathcal{A}_1$. Indeed, $\exp(t\, ad^\hbar_{a})=\exp(t\, ad_{a_1})$
is an automorphism of the $\hbar$-filtered associative algebra $\mathcal{A}\,\llangle\hbar\rrangle$ since $\exp(t\, ad_{a_1})$ is, by assumption, an automorphism of the almost-commutative algebra $\mathcal A$ (\textit{i.e.} it preserves the filtration).
 
Secondly, one can check by explicit computation that the lemma also holds for any element $b(\hbar)\in\mathcal{A}\,\llangle\hbar\rrangle\cap\hbar^2\,\mathcal{A}\llbracket\hbar\rrbracket$, that is to say an element of the form $b(\hbar)=\sum^\infty_{n=2} b_n\,\hbar^n$ with $b_n\in\mathcal{A}_n$:
\ba
\exp(t\, ad^\hbar_{b})&=&\exp\Big(\,t\sum^\infty_{n=2} \hbar^{n-1}\, ad_{b_n}\Big)\label{expadhbar}\\
&=&\sum\limits_{k=0}^\infty\frac{t^k}{k!}\sum\limits_{n_1,\cdots,\,n_k\geqslant 2} \hbar^{n_1+\cdots+n_k-k}\,ad_{{}_{b_{_{n_1}}}}\cdots ad_{{}_{b_{_{n_k}}}}
\ea
The main point to observe is that, at each order in $\hbar$, there will only be only a finite sum of products $ad_{{}_{b_{_{n_1}}}}\cdots ad_{{}_{b_{_{n_m}}}}$ (since each $n_i$ contributes to a strictly positive number in the exponent). This ensures that the exponential \eqref{expadhbar} is a well-defined $\hbar\,$-linear map on $\mathcal{A}\,\llangle\hbar\rrangle$\,. Moreover, since it is an exponential of a derivation of $\mathcal{A}\,\llangle\hbar\rrangle$, it is automatically an algebra automorphism.

Thirdly, the composition product of two automorphisms $\exp(ad^\hbar_{a})$ and $\exp(ad^\hbar_{b})$, with $a(\hbar)=a_1\hbar$ and $b(\hbar)=\sum^\infty_{n=2} b_n\,\hbar^n$ as above, is well-defined (since each of these two automorphisms is well-defined). 
Therefore, one should only check that this product takes the form $\exp(ad^\hbar_{c})$ for some element $c$ in $\mathcal{A}\,\llangle\hbar\rrangle$\,. The Baker–Campbell–Hausdorff formula guarantees that the element $c$ belongs to $\mathcal{A}\,\llbracket\hbar\rrbracket$ and is well-defined (since one knows that the above composition product is well-defined). A closer look at each term in its expansion as nested commutators ensures that $c\in\mathcal{A}\,\llangle\hbar\rrangle$\,.
In fact, it is well-known that $\exp(ad_T)=Ad_{{}_{\exp(T)}}$ for any endomorphism $T$, hence the operation $\exp(ad^\hbar_{a})\circ\exp(ad^\hbar_{b})=\exp(ad^\hbar_{c})$ is equivalent to a conjugation $Ad_g$ by the element $g=\exp(\tfrac1{\hbar}\,a)\exp(\tfrac1{\hbar}\,b)=\exp(\tfrac1{\hbar}\,c)$, where the Dynkin version of the Baker–Campbell–Hausdorff formula gives
\ba
&c&=\,\,\hbar\,\log\Big(\,\exp(\tfrac1{\hbar}\,a)\exp(\tfrac1{\hbar}\,b)\,\Big)\\
&&=a+b-\sum\limits_{n=2}^\infty\frac{(-1)^n}{n}
\sum\limits_{r_i+s_i>0}\underbrace{k(r_1,s_1,\ldots,r_n,s_n)}_{\in\mathbb{R}}\,\underbrace{(ad^\hbar_{a})^{r_1}(ad^\hbar_{b})^{s_1}\cdots(ad^\hbar_{a})^{r_n}(ad^\hbar_{b})^{s_n-1}b}_{=:\,\,c(r_1,s_1,\ldots,r_n,s_n)}\,.\label{BCH}
\ea
The explicit expression of the coefficients $k(r_1,s_1,\cdots,r_n,s_n)$ in \eqref{BCH} is well-known (see \textit{e.g.} the book \cite{BCHD}, p.117) but is not necessary for the proof. It is enough to check that each term $c(r_1,s_1,\cdots,r_n,s_n)$ belongs to $\mathcal{A}\,\llangle\hbar\rrangle$\,. This is true because, for any almost-commutative algebra $\mathcal{A}$, one has that $ad^\hbar_{\alpha}\beta\in\mathcal{A}\,\llangle\hbar\rrangle$ for any $\alpha,\beta\in\mathcal{A}\,\llangle\hbar\rrangle$ (see Subsection \ref{formalextalmcom}).

Fourthly, the Zassenhaus formula gives
\be
\exp\Big(\,\tfrac1{\hbar}\,(a+b)\,\Big)\,=\,\underbrace{\exp(\tfrac1{\hbar}\,a)\exp(\tfrac1{\hbar}\,b)}_{=\exp\big(\tfrac1{\hbar}\,c\big)}\exp(\tfrac1{\hbar}\,d_1)\exp(\tfrac1{\hbar}\,d_2)\cdots\,,
\ee
where the expressions of $d_i$ in terms of nested commutators of $a$ and $b$ is known recursively
(see \textit{e.g.} footnote 14 in \cite{BCHD}, p.365). Their qualitative form is enough to check that $d_i\in\mathcal{A}\,\llangle\hbar\rrangle\cap\hbar^2\,\mathcal{A}\llbracket\hbar\rrbracket$ (for $a(\hbar)$ and $b(\hbar)$ as above). Therefore, the second and third step of the proof guarantee that each factor in the product
\be
\exp(ad^\hbar_{a+b})\,=\,\exp(ad^\hbar_{c})\circ\exp(ad^\hbar_{d_1})\circ\exp(ad^\hbar_{d_2})\circ\cdots
\ee
is an automorphism of $\mathcal{A}\,\llangle\hbar\rrangle$\,.
This ends the proof.


\end{document}